\documentclass[a4paper,12pt]{article}
\usepackage[section]{placeins}  
\usepackage[dvipdfmx]{graphicx}
\usepackage{amsfonts,amsthm,amsmath,amssymb,graphicx,float,mathtools}
\usepackage{cite}
\usepackage{ulem}
\usepackage{xcolor}
\numberwithin{equation}{section}
\usepackage{color}
\usepackage{subcaption}
\usepackage{appendix}
\usepackage{bbm} 
\usepackage{tensor} 
\usepackage{verbatim} 
\usepackage[
	  pagebackref=false,
	  colorlinks=true,
      linkcolor=blue,
      urlcolor=blue,
      filecolor=black,
      citecolor=red,
      pdfstartview=FitV,
      pdftitle={},
        pdfauthor={},
        pdfsubject={},
        pdfkeywords={},
        pdfpagemode=None,
        bookmarksopen=true
      ]{hyperref}
\usepackage{ascmac}
\usepackage{tikz}
\usepackage{graphicx} 
\graphicspath{ {Pic/} }
\usepackage{subcaption}

\begin{document}

\newtheorem{definition}{Definition}[section]
\newcommand{\be}{\begin{equation}}
\newcommand{\ee}{\end{equation}}
\newcommand{\bea}{\begin{eqnarray}}
\newcommand{\eea}{\end{eqnarray}}
\newcommand{\CB}[1]{{\color{orange}{{\bf CB:} #1}}}
\newcommand{\XW}[1]{{\color{cyan}{{\bf XW:} #1}}}

\newcommand{\LE}{\left[}
\newcommand{\R}{\right]}
\newcommand{\nn}{\nonumber}
\newcommand{\Tr}{\text{Tr}}
\newcommand{\N}{\mathcal{N}}
\newcommand{\G}{\Gamma}
\newcommand{\vf}{\varphi}
\newcommand{\LL}{\mathcal{L}}
\newcommand{\Op}{\mathcal{O}}
\newcommand{\HH}{\mathcal{H}}
\newcommand{\arctanh}{\text{arctanh}}
\newcommand{\up}{\uparrow}
\newcommand{\down}{\downarrow}
\newcommand{\ket}[1]{\left| #1 \right>}
\newcommand{\bra}[1]{\left< #1 \right|}
\newcommand{\ketbra}[1]{\left|#1\right>\left<#1\right|}
\newcommand{\rd}{\partial}
\newcommand{\de}{\partial}
\newcommand{\ba}{\begin{eqnarray}}
\newcommand{\ea}{\end{eqnarray}}
\newcommand{\db}{\bar{\partial}}
\newcommand{\we}{\wedge}

\newcommand{\ca}{\mathcal}
\newcommand{\lr}{\leftrightarrow}
\newcommand{\f}{\frac}
\newcommand{\s}{\sqrt}
\newcommand{\vp}{\varphi}
\newcommand{\hvp}{\hat{\varphi}}
\newcommand{\tvp}{\tilde{\varphi}}
\newcommand{\tp}{\tilde{\phi}}
\newcommand{\ti}{\tilde}
\newcommand{\ap}{\alpha}
\newcommand{\pr}{\propto}
\newcommand{\mb}{\mathbf}
\newcommand{\ddd}{\cdot\cdot\cdot}
\newcommand{\no}{\nonumber \\}
\newcommand{\la}{\langle}
\newcommand{\lb}{\rangle}
\newcommand{\ep}{\epsilon}
 \def\we{\wedge}
 \def\lr{\leftrightarrow}
 \def\f {\frac}
 \def\ti{\tilde}
 \def\ap{\alpha}
 \def\pr{\propto}
 \def\mb{\mathbf}
 \def\ddd{\cdot\cdot\cdot}
 \def\no{\nonumber \\}
 \def\la{\langle}
 \def\lb{\rangle}
 \def\ep{\epsilon}
\newcommand{\mcl}{\mathcal}
 \def\g{\gamma}
\def\Tr{\text{tr}}

\begin{titlepage}
\thispagestyle{empty}

\begin{flushright}

\end{flushright}
\bigskip
\begin{center}
  \noindent{\large \textbf{Relaxation Process During Complex Time Evolution In Two-Dimensional Integrable and Chaotic CFTs}}
\vspace{2cm}

\vspace{1cm}
\renewcommand\thefootnote{\mbox{$\fnsymbol{footnote}$}}
Chen Bai \footnote{baichen22@mails.ucas.ac.cn}${}^{1}$,
Weibo Mao \footnote{maoweibo21@mails.ucas.ac.cn}${}^{1}$, 
Masahiro Nozaki\footnote{mnozaki@ucas.ac.cn}${}^{1,2}$,
Mao Tian Tan\footnote{mao.tan@fmf.uni-lj.si}${}^{3,4}$, and
Xueda Wen\footnote{xueda.wen@physics.gatech.edu}${}^{5}$
\\

\vspace{1cm}
${}^{1}${\small \sl Kavli Institute for Theoretical Sciences, University of Chinese Academy of Sciences,
Beijing 100190, China}\\
${}^{2}${\small \sl RIKEN Interdisciplinary Theoretical and Mathematical Sciences (iTHEMS), \\Wako, Saitama 351-0198, Japan}\\
${}^{3}${\small \sl Asia Pacific Center for Theoretical Physics, Pohang, Gyeongbuk, 37673, Korea}\\
${}^{4}${\small \sl Department of Physics, Faculty of Mathematics and Physics, University of Ljubljana, Jadranska 21, SI-1000 Ljubljana, Slovenia}\\
${}^{5}$ {\small \sl School of Physics, Georgia Institute of Technology, Atlanta, GA 30332, USA}\\
\vspace{1cm}
\end{center}
\begin{abstract}
	We investigate the complex time evolution of a vacuum state with the insertion of a local primary operator in two-dimensional conformal field theories (2d CFTs). 
    This complex time evolution can be considered as a composite process constructed from Lorentzian time evolution and a Euclidean evolution induced by a post-selected measurement.
    Our main finding is that in the spatially-compact system, this complex time evolution drives the state of the subsystems 
    to those of the primary state with the same conformal dimensions of the inserted operator. 
    Contrary to the compact system, the subsystems of the spatially non-compact system evolve to states that depend on the non-unitary process during a certain time regime. 
    In holographic systems with a compact spatial direction, this process induced by a heavy local operator can correspond to the relaxation from a black hole with an inhomogeneous horizon to that with a uniform one, while in the ones with a non-compact spatial direction, it can correspond to the relaxation to that with a horizon depending on the non-unitary process.

\end{abstract}
\end{titlepage} 
\tableofcontents

\section{Introduction \label{sec:introduction}}
Non-equilibrium physics is currently a central research topic across a wide range of fields, including high-energy physics, condensed matter physics, and quantum information.
One of the central non-equilibrium phenomena actively investigated in theoretical physics is the mechanism by which thermal states emerge from a quantum state. 
A crucial piece of the puzzle is to understand how subsystems thermalize.
The research subjects relevant to (subsystem) thermalization are information scrambling and quantum chaos, and these have been actively investigated in \cite{Hayden:2007cs,Sekino:2008he,Shenker:2013pqa,Maldacena:2015waa,Nahum:2017yvy,Lewis-Swan:2019xbi,Xu:2022vko,Schuster:2021uvg,Parker:2018yvk,Vermersch:2018sru,Garttner:2016mqj,Deutsch:1991msp,Srednicki:1994mfb,Popescu:2006rhr,Rigol:2007juv,DAlessio:2015qtq,Goldstein:2005aib,Tasaki:2016msp,Nandkishore:2014kca,Kaufman:2016mif,Neill:2016koy,Dymarsky:2016ntg,Lashkari:2016vgj,Berry:1977qtq,Bohigas:1983er,Maldacena:2016hyu,Oganesyan:2007wpd,Cotler:2016fpe,Fisher:2022qey,Stanford:2019vob,Chen:2024oqv,Gharibyan:2018jrp,Chang:2024lxt,Onoda:2025yue,Sugimoto:2023llh,Sugimoto:2021aao,Caputa:2024gve,Kawamoto:2024vzd,Chen:2024imd,Chen:2024lji,Nie:2018dfe,Kudler-Flam2020,PhysRevResearch.3.033182,PhysRevB.104.214202,Goto2022}.
In particular, in the AdS/CFT correspondence \cite{Maldacena:1997re,1998hep.th....3131W,Gubser:1998bc}, non-equilibrium phenomena are expected to be closely related to remarkable spacetime dynamics, such as the emergence of spacetime \cite{2010JHEP...11..054T,Hartman:2013qma,2015JHEP...05..152M}, and the growth of the black hole interior \cite{2016PhRvD..93h6006B,2014arXiv1411.0690S,2014arXiv1403.5695S,2015arXiv150907876B}.
    While unitary non-equilibrium dynamics has been extensively explored, it is natural to generalize these studies to non-unitary time evolution. Such extensions may provide new perspectives on black-hole formation, evaporation, and relaxation processes.

\medskip
Motivated by this perspective, there has been a growing interest in non-unitary time evolution spanning multiple fields of physics, 
including quantum field theory (QFT), quantum gravity, condensed matter physics and quantum information science. 

In QFT and quantum gravity, this growing attention is driven in part by developments in the formulation of quantum theories on complex spacetime geometries.
In ~\cite{Segal_2021}, Kontsevich and Segal (KS) investigated the class of complexified Lorentzian metrics for which the path-integral of a generic quantum field theory is manifestly convergent. They identified a broad, contractible space of such allowable metrics, which continuously deforms to the space of Euclidean metrics, providing a mathematically controlled framework for studying quantum field theories.

Building on this, Witten proposed to examine the KS condition in concrete examples, particularly in the context of quantum gravity~\cite{Witten2021_ComplexMetric}. He pointed out that many useful complex geometries, such as complexified black hole solutions, naturally satisfy the KS criterion, while certain pathological spacetimes, like complex wormholes, do not. These proposals have stimulated a wide range of investigations into the physical implications of complex spacetime metrics across diverse settings, including cosmology, gravitational thermodynamics, and holography~\cite{2022_Bond,Lehners2021,2022Cosmology,2022_Visser,2203_Loges,2206_Bris,2023_Hertog}. 
One simple class of allowable metrics introduced in~\cite{Witten2021_ComplexMetric} is
\begin{equation}
\label{Complex_metrics}
ds_{\pm}^2 = -(1 \mp i\delta)^2 dt^2 + d\vec{x}^{\,2}, \quad \delta > 0,
\end{equation}
which generates a non-unitary time evolution of the form
\begin{equation}
\label{complex_time_general}
|\psi(t)\rangle = e^{-iHt - \delta Ht} |\psi_0\rangle, \qquad \langle \psi(t)| = \langle \psi_0| e^{iHt - \delta Ht},
\end{equation}
with \( H \) denoting the system Hamiltonian. These evolutions include a damping term that tends to project the state toward the ground state of \( H \), resulting in an explicitly non-unitary time evolution. Some exactly solvable setups on such non-unitary time evolutions were recently studied in \cite{2024Nozaki,Wen_2024}, and it was found that there are interesting universal features in the time evolution of quantum entanglement. 

In parallel, non-unitary dynamics has also attracted significant interest in open quantum systems, where deviations from unitarity arise due to measurement processes or couplings to an external environment. Prominent examples include measurement-induced phase transitions~\cite{2019Skinner,2018_Li,1908_Bao,1908_Jian} and dissipative protocols for preparing long-range entangled states~\cite{2112_Nat,2206_Hsieh,2209_Nat,2208_Zhu}. It has been increasingly recognized that quantum measurements in many-body systems can be fruitfully interpreted as generating non-unitary evolution, a perspective that has led to a number of recent advances\cite{2023Altaman2,2023Altman,2023Jian,2023Jian2,2023Alicea,2024Alicea,2022Granet,2020Chen,2023Schiro,2021_Tang,Lapierre_2025_cft,Lapierre_2025_circuit,2025_Tang}.

\subsection*{Summary}
 
In this paper, we investigate the complex time evolution from the vacuum state with the insertion of a local operator by studying the energy density (see \ref{sec:Energy-density}),  second R\'enyi entanglement entropy (see Section \ref{sec:SREE}), and entanglement entropy (see Sections \ref{sec:2d-com-hol} and \ref{sec:2dhCFT-inf}).
 This complex time evolution can be considered as a composite process constructed out of Lorentzian time evolution and a non-unitary process induced by post-selected measurement.
In this paper, we considered the spatially compact and non-compact systems that can be described by the two-dimensional free and holographic conformal field theories ($2$d free and holographic CFTs).
Here, the $2$d holographic CFTs are those that admit dual gravitational description.

We found that in the compact systems under consideration, the complex time evolution evolves from the reduced density matrix 
of the vacuum state with an insertion of a primary operator to that of a primary state with the same conformal dimensions as those of the inserted operator.
On the other hand,
during a certain time region of the complex time evolution, the 
reduced density matrices
of the $2$d holographic system with a spatially non-compact direction evolve to ones that depend on the non-unitary process.
In particular, in the large time regime of the complex time evolution, the semi-infinite subsystem saturates to a state that depends on the non-unitary process.

Furthermore, we also investigated the gravitational dual of systems with an insertion of a heavy operator (see Section \ref{Sec:Gravity-Duals}).
The gravity dual is initially described as a black hole with an inhomogeneous horizon.
In the compact system, the black hole with an inhomogeneous horizon relaxes to that with a homogeneous one, while in the non-compact system, it relaxes to a horizon that depends on the non-unitary process.
On the CFT side, this can be considered as the relaxation to a thermal state that depends on the conformal dimensions of the inserted primary operator.

\subsection*{Organization of this paper}
The organization of this paper is as follows.

In Section \ref{Sec:Local-Operator-Quenches}, we will give an interpretation of the complex time evolution as a non-unitary process induced by post-selected measurement and then introduce the definition of (R\'enyi) entanglement entropy.
Furthermore, we will investigate the non-equilibrium phenomena during the time evolution from the vacuum state with an insertion of a local operator by using the expectation of the energy density and entanglement entropy.
	We will consider the systems described by free CFTs and holographic CFTs, i.e., those described by gravity. 
	Then, we will report our findings, e.g., the relaxation from the vacuum state with an insertion of a primary operator to a primary state with the same conformal dimensions as those of the inserted operator.
	
	In Section \ref{Sec:Gravity-Duals}, we will propose the gravity dual that may reflect the properties of the non-unitary time evolution under consideration, investigate the remarkable properties of this gravity dual, and report and give some interpretations on our findings.
	
	In Section \ref{Sec:Discussions-and-Future-Directions}, we will discuss our findings and comment on future directions.

    Note that in this paper, all parameters and variables are dimensionless. 
    The unit of the length scale is assumed to be a lattice spacing which will be introduced later to make the quantities considered in this paper finite.
	
	
\section{Local Operator Quenches with Complex Time}
\label{Sec:Local-Operator-Quenches}

In this section, we will investigate entanglement dynamics induced by a complex time local operator quench. 
The systems under consideration can be described by $2$d CFTs.
In this paper, we consider both spatially compact systems with circumference $L$ as well as spatially infinite systems.
We start from $\ket{0}$, the vacuum state of the $2$d uniform Hamiltonian $H$, with an insertion, at spatial point $x$, of a local primary operator with conformal dimensions $(h_{\mathcal O},\overline{h}_{\mathcal O})$ ,
and then evolve it with $H$ for a complex time interval of $t-i\delta t$ with $\delta\ge 0$.
Then, the system is in the state
\begin{equation}
\label{eq:system-with-LO}
\ket{\Psi_{\mathcal O}(t)}=\mathcal N_{\mathcal O}(t)\;e^{-iH(t-i\delta t)}\,e^{-\beta H}\,\mathcal O(0,x)\ket{0}\,,
\end{equation}
where $\beta>0$ is a short-distance regulator that tames the divergence induced by the contraction of the local operators.
To preserve the norm of this state at unity, choose the normalization factor, $\mathcal N_{\mathcal O}(t)$, as
\begin{equation}
\label{eq:N-def}
\mathcal N_{\mathcal O}(t)^{-2}=\bra{0}\mathcal O^\dagger(x)\,e^{-\,2(\beta+\delta t)\,H}\,\mathcal O(x)\ket{0}\,.
\end{equation}
The density matrix is  
\begin{equation}
\label{eq:density-matrix}
\rho(t)=\ket{{\Psi_{\mathcal O}(t)}}\bra{{\Psi_{\mathcal O}(t)}}=\mathcal N_{\mathcal O}(t)^{2}\;e^{-iHt}\,e^{-(\beta+\delta t)H}\,\mathcal O(x)\ket{0}\bra{0}\,\mathcal O^\dagger(x)\,e^{-(\beta+\delta t)H}\,e^{iHt}.
\end{equation}
\subsection{Euclidean Time Evolution from Projective Measurement \label{eq:Interpretation-as-measurement}}
This complex time evolution can be considered as a process on a geometry described by the complex metric,
\be \label{eq:com-geo-und-con}
ds^2=-(1-i\delta)^2dt^2+dx^2, ~~\delta >0
\ee
where $t$ and $x$ are the coordinates for the temporal and spatial directions, respectively.
Here, we provide another interpretation of the complex time evolution as the composite time evolution constructed out of Lorentzian time evolution and the Euclidean time evolution induced by post-selected measurement on the ancilla \cite{2025PhRvB.112j4322L}.

 Extend our system to $\mathcal{C}$, a composite system that consists of the system under consideration, $\mathcal{S}$, and $n$ ancillas, each of which is assumed to be a two-level system for simplicity.
Furthermore, suppose that we start from a tensor product state which is defined as
\be
\rho_{\mathcal{C}}= \f{\rho_{I}\prod_{j=1}^{m}\otimes\rho_{\text{an};j}}{\bra{0}\mathcal O^\dagger(x)\,e^{-\,2\beta\,H}\,\mathcal O(x)\ket{0}},
\ee
where $\rho_{\text{an};i}$ denotes the density operator associated with $i$-th ancilla, while $\rho_{I}$ is defined as 
\be
\rho_{I}= e^{-\beta H}\mathcal{O}(x) \ket{0}\bra{0}\mathcal{O}^{\dagger}(x)e^{-\beta H}.
\ee
In other words, $\rho_{I}$ is the initial density operator under consideration, i.e., (\ref{eq:density-matrix}) with $t=0$.
For simplicity, we assume that the density operator associated with each ancilla is 
\be
\rho_{\text{an};i}=\ket{+}_i\bra{+}_i,
\ee
where $\ket{+}_i$ is the eigenvector of $\sigma_z^i$, the $\sigma_z$ Pauli matrix acting on the i-th ancilla, with eigenvalue $+1$.

We take the interaction between $\mathcal{S}$ and the i-th ancilla to be described by the following unitary

\be \label{eq:Interaction-between-ancilla-system}
U_{\mathcal{S};i}=e^{-i \mu \mathcal{J}_{\mathcal{S}} \sigma^i_{x}},
\ee
where $\mathcal{J}_{\mathcal{S}}$ is a Hermitian operator acting only on $\mathcal{S}$, while $\sigma^i_{x}$ is the $\sigma_x$ Pauli matrix acting on the $i$-th ancilla.
The action of this operator acting on $i$-th ancilla is $\sigma^i_{x}\ket{\pm}_i=\ket{\mp}_i$, where $\sigma^i_{z}\ket{\pm}_i=\pm\ket{\pm}_i$.
Here, $\mu$ is the duration during which $\mathcal{S}$ and the $i$-th ancilla are allowed to interact.
It should be pointed out that we can express (\ref{eq:Interaction-between-ancilla-system}) as
\be \label{eq:BH-formula}
U_{\mathcal{S};i}=\underset{=\sum_{l=0}^{\infty}\f{(-i \mu \mathcal{J}_{\mathcal{S}})^{2l}}{(2l)!}}{\underbrace{\cos{\left(\mu \mathcal{J}_{\mathcal{S}}\right)}}\cdot{\bf 1}_{\text{an}}}-(i\mu)\sum_{l=0}^{\infty} \f{(-\mu^2)^l}{(2l+1)!} \mathcal{J}_{\mathcal{S}}^{2l+1}\cdot\sigma^i_x.
\ee
Let us evolve the composite system $\mathcal{C}$ with $U_{\mathcal{S};i}$, followed by $e^{-iH \Delta t}$, where $\Delta t$ is a positive parameter. 
After acting on the composite system with this sequence of unitary operators, we perform post-selected measurement on the $i$-th ancilla with $\ket{+}_{i}$. 
As a consequence, the system is in the state
\be \label{eq:projected-op-on-i}
\begin{split}
    \bra{+}_i\rho'_{\mathcal{C}}\ket{+}_i= \bra{+}_ie^{-iH\Delta t}U_{\mathcal{S};i} \rho_{\mathcal{C}} U^{\dagger}_{\mathcal{S};i}e^{iH \Delta t}\ket{+}_i=e^{-iH\Delta t}\cos{\left(\mu \mathcal{J}_{\mathcal{S}}\right)} \rho_{\mathcal{C}/i} \cos{\left(\mu \mathcal{J}_{\mathcal{S}}\right)}e^{iH\Delta t},
\end{split}
\ee
where in the last equation, we exploited (\ref{eq:BH-formula}). Here, $\rho_{\mathcal{C}/i}$ is defined as
\be
\rho_{\mathcal{C}/i}= \rho_{I}\prod_{j\neq i}^{m}\otimes\rho_{\text{an};j}.
\ee
Assume that we can find $\mu \mathcal{J}_{\mathcal{S}}$ such that
\be \label{eq:equation-for-muA}
\cos{\left(\mu \mathcal{J}_{\mathcal{S}}\right)} =e^{-H\delta \Delta t},
\ee
Then, we can consider this Euclidean time evolution as a non-unitary time evolution induced by the post-selected measurement.
Here, we assume that we can find $\mu \mathcal{J}_{\mathcal{S}}$ satisfying (\ref{eq:equation-for-muA}).
We define the projected density operator as
\be
\tilde{\rho}_{\mathcal{C}/i}=\f{e^{-iH\Delta t}e^{-H\delta \Delta t} \rho_{\mathcal{C}/i} e^{-H\delta \Delta t} e^{iH \Delta t}}{\bra{0}\mathcal O^\dagger(x)e^{-\beta  H}e^{-2H\delta \Delta t} e^{-\beta  H}\mathcal O(x)\ket{0}}.
\ee
Note that (\ref{eq:projected-op-on-i}) possesses three properties: (1) $\Tr \tilde{\rho}_{\mathcal{C}/i}=1$; (2) $\tilde{\rho}_{\mathcal{C}/i}^{\dagger}=\tilde{\rho}_{\mathcal{C}/i}$; and (3) all the eigenvalues of $\tilde{\rho}_{\mathcal{C}/i}$ are positive.
Then, we define $M$, a single-step process, as 
\be \label{eq:Sigle-step-process}
M=e^{-iH\Delta t}e^{-\delta H \Delta t},
\ee
where $\delta$ can be thought of as the parameter which determines how strongly the non-unitary process induced by the post-selected measurement influences the system compared to the Lorentzian time evolution. 

After repeating this process with every ancilla as in (\ref{eq:projected-op-on-i}), the density matrix after all post-selected measurements is
\be \label{eq:measuremented-density-op}
\begin{split}
\rho_{\text{A.M}}
    &= \f{M^m \rho_{I} M^{\dagger m}}{\bra{0}\mathcal O^\dagger(x)\,e^{-\,2(\beta+\delta m \Delta t)\,H}\,\mathcal O(x)\ket{0}}\\
&=\f{e^{-i H m \Delta t}e^{-H m\Delta t} e^{-\beta H}\mathcal{O}(x)\ket{0}\bra{0} \mathcal{O}^{\dagger}(x)e^{-\beta H}e^{-H m \Delta t}e^{-i H m \Delta t}}{\bra{0}\mathcal O^\dagger(x)\,e^{-\,2(\beta+\delta m \Delta t) \,H}\,\mathcal O(x)\ket{0}}.
\end{split}
\ee
In this section, we implicitly treat $m$ as an integer. 
By setting $m\Delta t =t $, we can identify (\ref{eq:measuremented-density-op}) with (\ref{eq:density-matrix}).

\subsection{Definition of Entanglement Entropy \label{sec:D-EE}}
In this paper, we investigate the properties of the system in (\ref{eq:system-with-LO}) by exploiting (R\'enyi) entanglement entropy and the expectation value of the energy density.

In this section, we review the definition of the (R\'enyi) entanglement entropy and the analytical 
computational method which uses the Euclidean-path integral.
To this end, we start from a system, and divide it into $A$ and $\overline{A}$, the complement of $A$.
Then, we define a reduced density matrix associated with $A$ as 
\be \label{eq:reduce-density-matrix}
\rho_A(t)=\Tr_{\overline{A}}\rho(t),
\ee
where the operation, $\Tr_{\over{A}}(\cdot)$, denotes the partial trace associated with  $\overline{A}$.
Subsequently, define entanglement entropy, 
the quantity characterizing the bipartite entanglement between $A$ and $\overline{A}$, as the von Neumann entropy for the reduced density matrix, i.e.,
\be
S_{A}(t) = -\Tr_A\left[\rho_A(t)\log{\rho_A(t)}\right].
\ee
Furthermore, the $n$-th R\'enyi entanglement entropy is defined as the one-parameter generalization of the entanglement entropy, i.e., 
\be
S^{(n)}_A(t)=\f{1}{1-n}\log{\left(\Tr_A\rho^n_A(t)\right)},
\ee
where in the von Neumann limit, $n\rightarrow 1$,  R\'enyi entanglement entropy reduces to the entanglement entropy as 
\be
\lim_{n\rightarrow 1} S^{(n)}_A(t)=S_A(t).
\ee

To analytically calculate these quantities, we start with the Euclidean path-integral. 
We define a Euclidean counterpart of (\ref{eq:density-matrix}) as 
\begin{equation}
\label{eq:euc-den-for-los}
\rho_{\rm E}(\tau_1,\tau_2)=\mathcal N_{\mathcal O}(\tau_1)^{2}\;e^{-(\beta+\delta\tau_1+\tau_2)H}\,\mathcal O(x)\ket{0}\bra{0}\,\mathcal O^\dagger(x)\,e^{-(\beta+\delta\tau_1-\tau_2)H},
\end{equation}
where $\tau_1,\tau_2\in\mathbb R$ and the normalization constant, $\mathcal N_{\mathcal O}(\tau_1)$, satisfies
\be
\mathcal N_{\mathcal O}^{-2}(\tau_1)=\bra{0}\mathcal O^\dagger(x)\,e^{-\,2(\beta+\delta \tau_1)\,H}\,\mathcal O(x)\ket{0}\,.
\ee
Then, we define the Euclidean part of (\ref{eq:reduce-density-matrix}) as
\be
\rho_{\rm E;A}(\tau_1,\tau_2)=\Tr_{\overline{A}}\,\rho_{\rm E}(\tau_1,\tau_2)
\ee
Subsequently, 
the Euclidean counterparts of entanglement entropy and R\'enyi entanglement entropy are
\be
\begin{split}
   & S_{A;E}=-\Tr_{A}\left[\rho_{\rm E;A}(\tau_1,\tau_2)\log{\rho_{\rm E;A}(\tau_1,\tau_2)}\right],\\
   & S^{(n)}_{A;E}=\f{1}{1-n}\log{\left(\Tr_A \rho^n_{\rm E;A}(\tau_1,\tau_2)\right)}.
\end{split}
\ee
From here on, we take $A$ to be a  spatial interval such that $A=[X_2,X_1]$ with $0<X_2<X_1$.
By employing the twist-operator formalism \cite{Calabrese_2004,2009JPhA...42X4005C}, we obtain the $n$-th moment of Euclidean R\'enyi entanglement entropy as 
\begin{equation}
\label{eq:Euc-REE-LOS}
S^{(n)}_{A;{\rm E}}=\frac{1}{1-n}\log\!\left[
\frac{\big\langle \mathcal O_n^\dagger(w_2,\overline{w}_2)\,\sigma_n(w_{X_1},\overline{w}_{X_1})\,\overline{\sigma}_n(w_{X_2},\overline{w}_{X_2})\,\mathcal O_n(w_1,\overline{w}_1)\big\rangle}
{\big\langle \mathcal O^\dagger(w_2,\overline{w}_2)\,\mathcal O(w_1,\overline{w}_1)\big\rangle^{\,n}}
\right],
\end{equation}
where the twist and anti-twist operators are primary with conformal dimensions $\left(\f{c(n^2-1)}{24n}, \f{c(n^2-1)}{24n}\right)$, while $\mathcal O_n$ is a primary operator with conformal dimension $\left(n h_{\mathcal O},n \overline{h}_{\mathcal O}\right)$.
Furthermore, we introduce the complex coordinates  $w=\tau+ix$, $\overline{w}=\tau-ix$.
In these coordinates, let $w_{i}$, $\overline{w}_{i}$, 
$w_{X_i}$, and $\overline{w}_{X_i}$ denote
\begin{equation}
\begin{split}
    \label{eq:insert-positions}
&w_{X_i}=iX_i,\quad w_1=-(\beta+\delta\tau_1+\tau_2)+ix,\quad \overline{w}_1=-(\beta+\delta\tau_1+\tau_2)-ix,\\
&\overline{w}_{X_i}=-iX_i,\quad w_2=(\beta+\delta\tau_1-\tau_2)+ix,\quad \overline{w}_2=-(\beta+\delta\tau_1-\tau_2)-ix.
\end{split}
\end{equation}

Ultimately, we will perform the analytic continuation $\tau_1=t$, $\tau_2=it$, and we will investigate the time dependence of the expectation value of the energy density and the (R\'enyi) entanglement entropy.

\if[0]
\textcolor{red}{{\bf I move the followings to other place:}
Mapping to the plane by $z=e^{\frac{2\pi}{L}w}$, $\overline{z}=e^{\frac{2\pi}{L}\overline{w}}$ yields
\begin{equation}
\label{eq:EE-cylinder-to-plane}
\begin{aligned}
S^{(n)}_{A;{\rm E}}&=\frac{1}{1-n}\log\!\left[
\frac{\big\langle \mathcal O_n^\dagger(z_2,\overline{z}_2)\,\sigma_n(z_{X_1},\overline{z}_{X_1})\,\overline{\sigma}_n(z_{X_2},\overline{z}_{X_2})\,\mathcal O_n(z_1,\overline{z}_1)\big\rangle_{\mathbb C}}
{\big\langle \mathcal O^\dagger(z_2,\overline{z}_2)\,\mathcal O(z_1,\overline{z}_1)\big\rangle_{\mathbb C}^{\,n}}
\right]\\
&\qquad -\frac{c(1+n)}{24n}\log\!\left[\prod_{i=1}^{2}\left|\frac{dz_{X_i}}{dw_{X_i}}\right|\right],
\end{aligned}
\end{equation}
where $c$ is the central charge.
The detailed Jacobians and uniformization are deferred to App.~\ref{app:uniformization}.
and we analytically continue by $\tau_1=t$, $\tau_2=it$.}
\fi
\subsection{Energy density}
\label{sec:Energy-density}

Before reporting the time dependence of the (R\'enyi) entanglement entropy, we report the time dependence of the expectation values of chiral and anti-chiral energy densities, which are universal in $2$d CFTs.
We define the expectation values of the chiral and anti-chiral energy-momentum operators, the quantities that can probe the local properties of the system, i.e., 
at a single spatial point.
For later convenience, we call the expectation values of the chiral and anti-chiral energy momentum operators chiral and anti-chiral energy densities.
Here, we begin with those energy densities in the compact system with spatial circumference of $L$.
To calculate the chiral and anti-chiral energy densities analytically, we employ the Euclidean path-integral.
Exploiting the conformal Ward–Takahashi identity on the plane and the map to the cylinder (for details, see App.~\ref{app:uniformization}), and then mapping the cylinder to the complex plane via $z = e^{\frac{2\pi}{L}w}$, $\overline{z} = e^{\frac{2\pi}{L}\overline{w}}$, we obtain the Euclidean expectation value of the chiral and anti-chiral energy densities in the system as
\begin{equation}
	\label{eq:ED-general}
	\begin{aligned}
		\big\langle T (w_X)\big\rangle
		&=-\frac{c}{24}\!\left(\frac{2\pi}{L}\right)^2+ h_{\mathcal O}\!\left(\frac{dz_X}{dw_X}\right)^{\!2}
		\left[\frac{1}{z_X-z_1}-\frac{1}{z_X-z_2}\right]^{\!2},\\
		\big\langle \overline{T} (\overline{w}_X)\big\rangle
		&=-\frac{c}{24}\!\left(\frac{2\pi}{L}\right)^2+ \overline{h}_{\mathcal O}\!\left(\frac{d\overline{z}_X}{d\overline{w}_X}\right)^{\!2}
		\left[\frac{1}{\overline{z}_X-\overline{z}_1}-\frac{1}{\overline{z}_X-\overline{z}_2}\right]^{\!2},2
	\end{aligned}
\end{equation}
where the first term is the vacuum energy on the cylinder, and 
$w_X = i X$, $\overline{w}_X = -i X$, 
$z_X = e^{\frac{2\pi}{L}w_X}$, and $\overline{z}_X = e^{\frac{2\pi}{L}\overline{w}_X}$.
Furthermore, $z_i$ and $\overline{z}_i$ denote the locations, on the plane, of the inserted local operators.

After performing the analytic continuation as $\tau_1=t$ and $\tau_2=it$, one obtains the time dependence of the chiral and anti-chiral energy densities as
\begin{equation}
	\label{eq:ED-analytic}
	\begin{aligned}
		\big\langle T (w_X)\big\rangle 
		&= -\frac{\pi ^2 \left(c-\frac{24 h_{\mathcal{O}} \sinh ^2\left(\frac{2 \pi  (\beta +t \delta )}{L}\right)}{\left(\cos \left(\frac{2 \pi  (-t+x-X)}{L}\right)-\cosh \left(\frac{2 \pi  (\beta +t \delta )}{L}\right)\right)^2}\right)}{6 L^2},\\
		\big\langle \overline{T} (\overline{w}_X)\big\rangle 
		&=-\frac{\pi ^2 \left(c-\frac{24 \overline{h}_{\mathcal{O}} \sinh ^2\left(\frac{2 \pi  (\beta +t \delta )}{L}\right)}{\left(\cos \left(\frac{2 \pi  (t+x-X)}{L}\right)-\cosh \left(\frac{2 \pi  (\beta +t \delta )}{L}\right)\right)^2}\right)}{6 L^2}.
	\end{aligned}
\end{equation}

\begin{figure}[htbp]
  \centering
  \subfloat[$\delta=0$]{\includegraphics[width=.35\columnwidth]{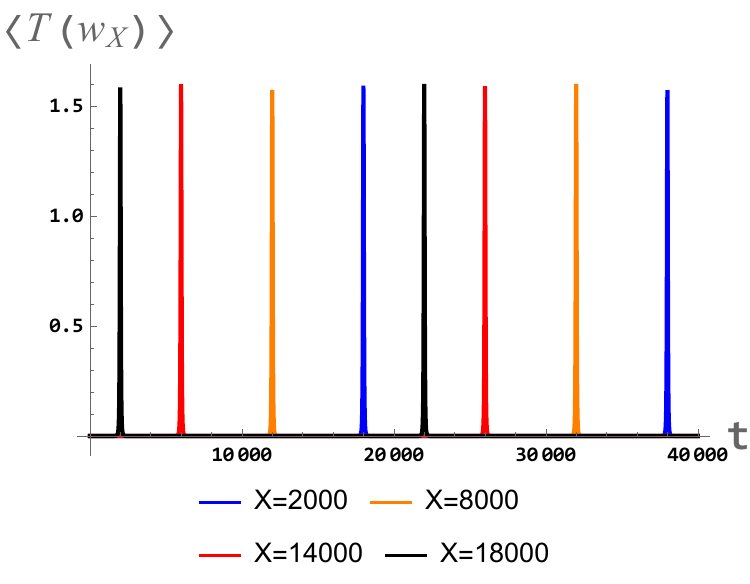}}\hspace{6pt}
  \subfloat[$\delta=0$]{\includegraphics[width=.35\columnwidth]{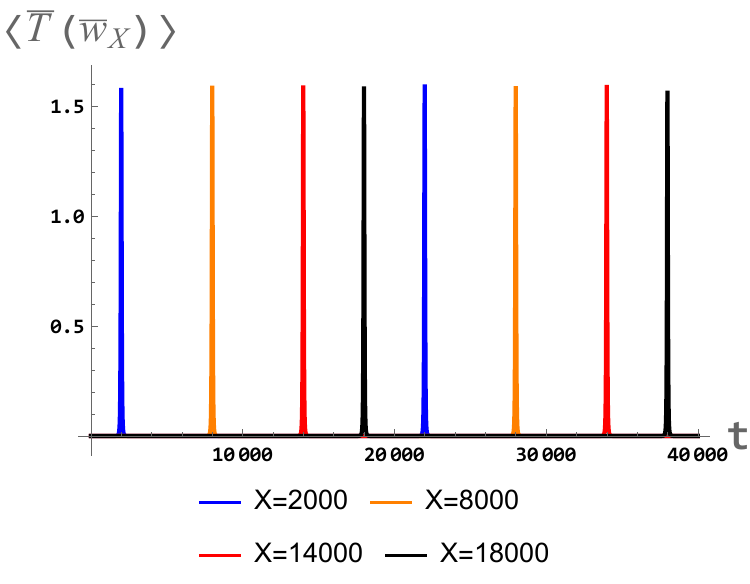}}\\
  \subfloat[$\delta=0.01$]{\includegraphics[width=.35\columnwidth]{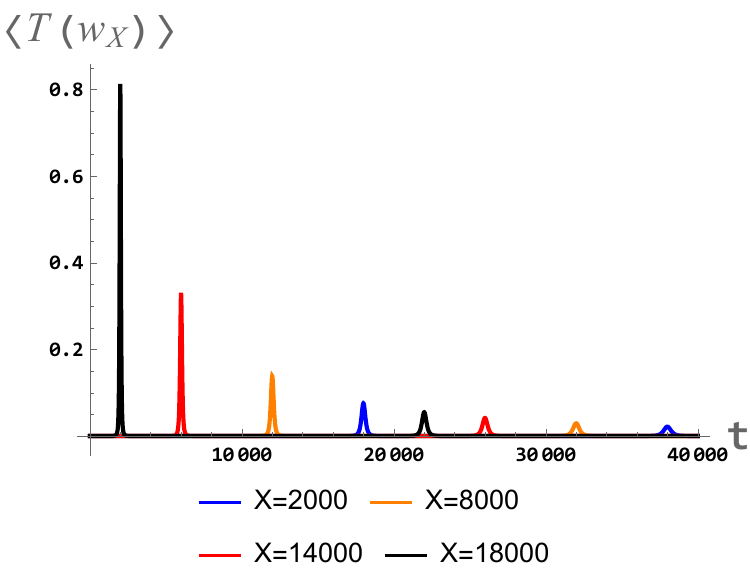}}\hspace{6pt}
  \subfloat[$\delta=0.01$]{\includegraphics[width=.35\columnwidth]{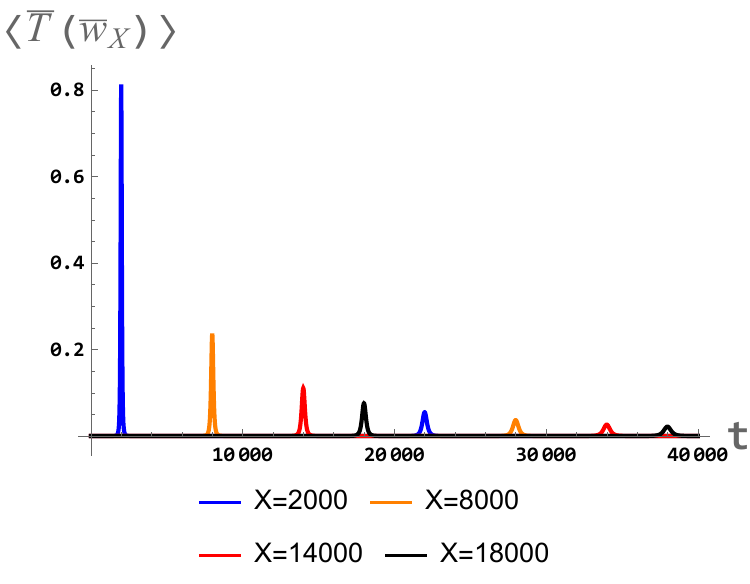}}\\
  \subfloat[$\delta=0.1$]{\includegraphics[width=.35\columnwidth]{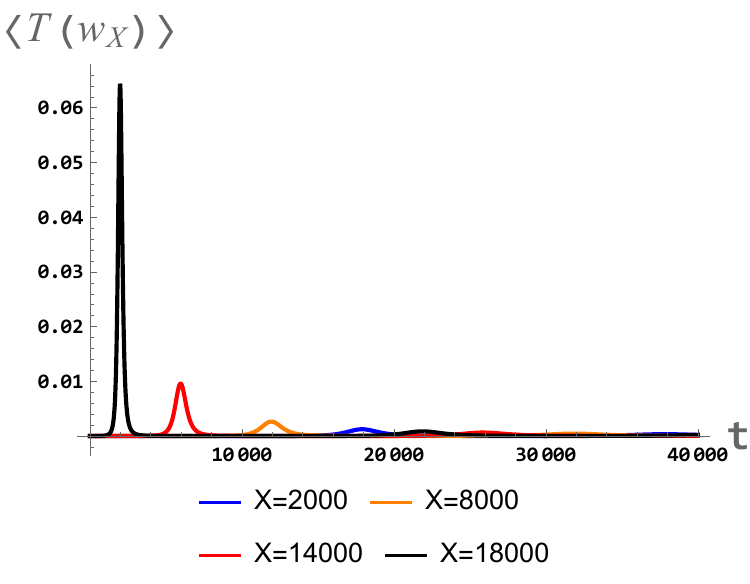}}\hspace{6pt}
  \subfloat[$\delta=0.1$]{\includegraphics[width=.35\columnwidth]{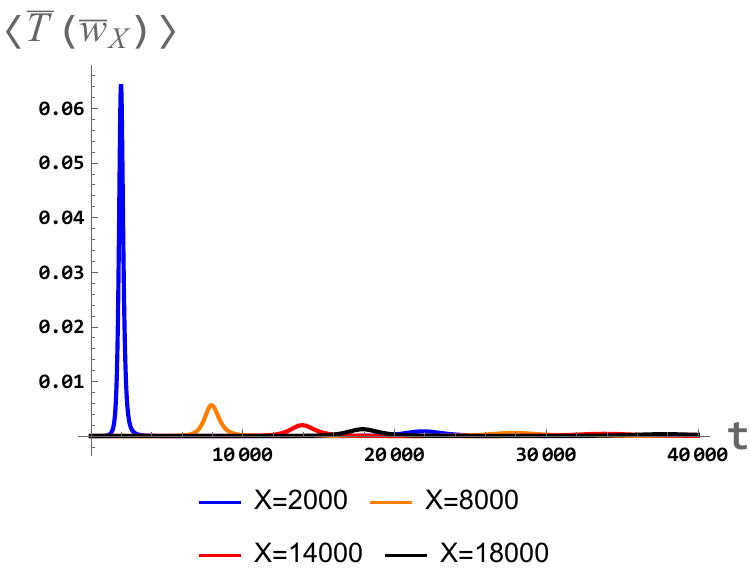}}\\
  \subfloat[$\delta=1$]{\includegraphics[width=.35\columnwidth]{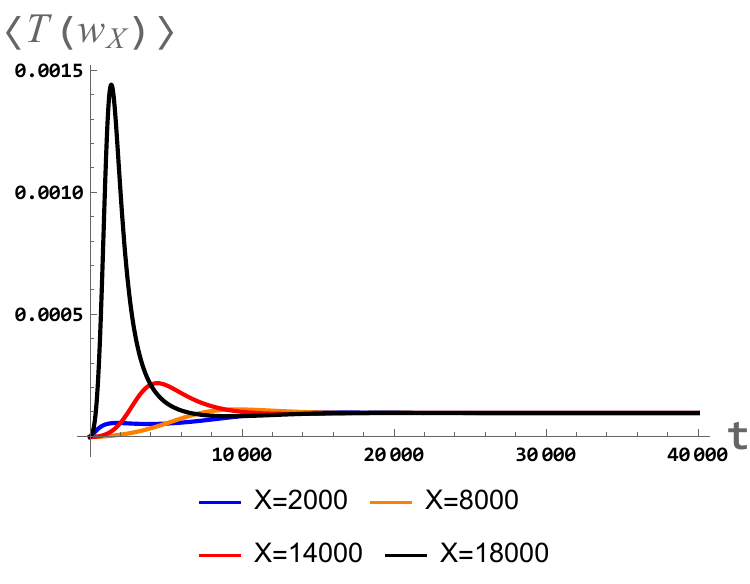}}\hspace{6pt}
  \subfloat[$\delta=1$]{\includegraphics[width=.35\columnwidth]{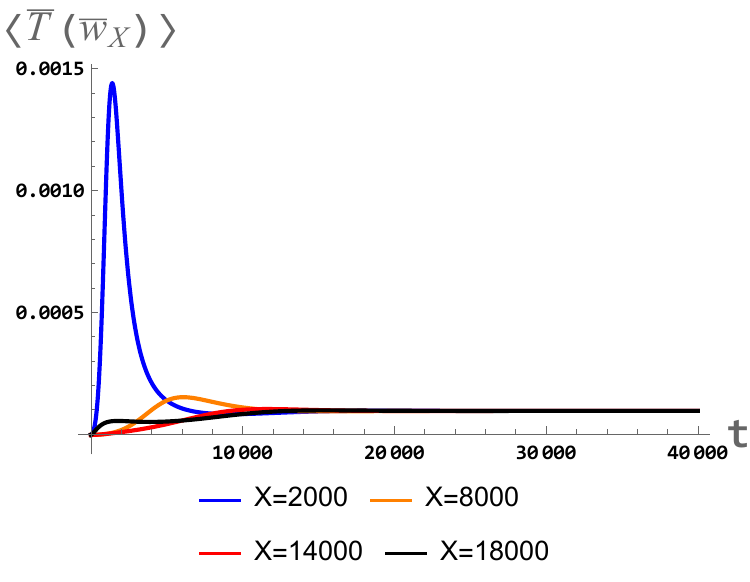}}
  \caption{ Chiral and anti-chiral energy densities for various $X$ and $\delta$, as a function of time. 
In the left panels, we show the time dependence of the chiral energy density, while in the right panels, we show that of the anti-chiral energy density.
  We set the parameters to be $\beta=50$, $x=0$ (insertion point), $L=20000$, $h_{\mathcal O}=\overline{h}_{\mathcal O}=1000$, $c=1000$.  
  }
  \label{Fig:ED-Circle}
\end{figure}

In Fig. \ref{Fig:ED-Circle}, we show the time dependence of the energy densities for various $\delta$ and $X$.
In this figure, the local operator is inserted at $x=0$. 
From this figure, we can observe that for $\delta=0$, i.e., the unitary time evolution, the time dependence of the energy densities exhibits  persistent periodic behavior with the period  $L$.
The time dependence of the energy densities exhibits energy peaks at certain times.
In the left panel with $\delta =0$, the energy peaks emerge around $t=nL- X$, where $n$ is an integer. On the other hand, in the right panel, they emerge around $t=nL+X$. This is because with $\delta=0$, the time dependence of (\ref{eq:ED-analytic}) is determined by $\cos{\left(\f{\pm t +x-X }{L}\right)}$. 

This periodic behavior can be described by the relativistic propagation of the two excitations induced by the local operator. 
One of the excitations moves to the left at the speed of light, while the other moves to the right.
When the left-moving excitation arrives at $X$, the chiral energy density exhibits an energy peak, and when the right one arrives at $X$, the anti-chiral one does.

Furthermore, from this figure, for $\delta =0.1$ and $0.01$, we can observe that although the chiral and anti-chiral energy density exhibits periodic behavior in time, according to time evolution, the height of energy-peaks decays, while the width becomes wider.
Furthermore, we observe that if $\delta$ becomes larger, the height of the energy peaks becomes smaller, and their width increases.
For $\delta =1$, we observe that the time dependence of the chiral and anti-chiral energy densities exhibits saturation to a certain value instead of persistent oscillatory behavior. 



Next, we focus on the late-time behavior of chiral and anti-chiral energy densities with $\delta \neq 0$.
By expanding the analytic expressions in ~\eqref{eq:ED-analytic} with respect to $\delta t \gg 1$, we can obtain the  late-time behavior as the leading term in this expansion,
\begin{equation} 
	\label{eq:ED-late}
	\begin{aligned}
	\big\langle T (w_{X})\big\rangle
	&\underset{\delta t\gg 1}{\approx}
	-\frac{\pi^2}{6L^2}\big(c-24h_{\mathcal O}\big),\\
	\big\langle \overline{T} (\overline{w}_{X})\big\rangle
	&\underset{\delta t\gg 1}{\approx}
	-\frac{\pi^2}{6L^2}\big(c-24\overline{h}_{\mathcal O}\big).
	\end{aligned}
\end{equation}

Thus, the expectation values of the chiral and anti-chiral energy densities, to leading order in a large time expansion, are equal to those of a primary state with the same conformal dimensions as the inserted local operator.
It should be pointed out that although the leading behavior in (\ref{eq:ED-late}) is independent of $\delta$ (
as long as $\delta >0$), this late-time behavior is valid.
Therefore, this late-time behavior can be interpreted as arising from a non-unitary evolution induced by the post-selected measurement performed on the ancillas.

Furthermore, let us interpret (\ref{eq:ED-late}) 
in terms of a  reduced density matrix.
The expectation values of the chiral and anti-chiral energy densities can be expressed as 
\be
\big\langle T_{tt}(X)\big\rangle = \Tr_{X} \left(T_{tt}(X) \rho_{X}(t)\right), 
\ee
where the energy density operator $T_{tt}(X)$ is defined as $T_{tt}(X)=T (w_{X})+\overline{T} (\overline{w}_{X})$ and $\rho_{X}$ is defined as $\rho_{X}=\Tr_{\overline{X}} \rho$, where $\overline{X}$ denotes the complement space to $X$, the insertion point of the energy density.
When we expand the density operator with respect to $t$, we assume that $\rho (t)$ can be expanded as
\be \label{eq:large-time-expansion-of-rdo}
\rho(t)=\rho^{(0)}(t)+\delta \rho(t),
\ee
where $\rho^{(0)}(t)$ denotes the leading term of $\rho(t)$ in this large $t$ expansion, while $\delta \rho(t)$ denotes the correction coming from higher order terms.
The late-time behavior in (\ref{eq:ED-late}) shows that $\rho^{(0)}_X(t)$ is time-independent and it is the reduced density operator of the primary state with the same conformal dimensions as those of the inserted operator, i.e., $\rho_{X}^{(0)}(t) = \Tr_{\overline{X}}\left(\ket{h_{\mathcal{O}},\overline{h}_{\mathcal{O}}}\bra{h_{\mathcal{O}},\overline{h}_{\mathcal{O}}}\right),
$ where $\ket{h_{\mathcal{O}},\overline{h}_{\mathcal{O}}}$ is the primary state with $(h_{\mathcal{O}},\overline{h}_{\mathcal{O}})$.
In other words, the system locally evolves, during complex time dynamics, to a primary state with the same conformal weights as the inserted operator.

Next, we present the time dependence of the chiral and anti-chiral energy densities in the spatially-infinite system, i.e., the system with $L\rightarrow\infty$.
By expanding (\ref{eq:ED-analytic}) with respect to $L$, we obtain the time dependence of the chiral and anti-chiral energy densities at the leading order in the large $L$ expansion as
\be \label{eq:EDS-inf}
\begin{split} 
    \big\langle T (w_{X})\big\rangle
    =\f{4h_{\mathcal{O}}(\beta +\delta t)^2}{\left[(-t+x-X)^2+(\beta +\delta t)^2\right]^2},~
    \big\langle \overline{T} (\overline{w}_{X})\big\rangle
    =\f{4\overline{h}_{\mathcal{O}}(\beta +\delta t)^2}{\left[(t+x-X)^2+(\beta +\delta t)^2\right]^2}.
\end{split}
\ee

The time dependence of the energy densities with $\delta=0$ is consistent with the findings of \cite{Caputa:2014vaa}.
In the large $t$ expansion, i.e., $\delta t \gg 1$, the leading behaviors of the chiral and anti-chiral energy densities are obtained as
\be \label{eq:late-EDS-inf}
\begin{split}
\big\langle T (w_{X})\big\rangle\underset{t\gg 1}{\approx} \begin{cases}
\f{4h_{\mathcal{O}}\beta^2}{t^4}&~\text{for}~\delta = 0 \\
\f{4h_{\mathcal{O}} \delta^2}{(1+\delta^2)^2 t^2} &~\text{for}~\delta \neq 0
\end{cases},\qquad
\big\langle \overline{T} (\overline{w}_{X})\big\rangle\underset{t\gg 1}{\approx} \begin{cases}
\f{4\overline{h}_{\mathcal{O}}\beta^2}{t^4}&~\text{for}~\delta = 0 \\
\f{4\overline{h}_{\mathcal{O}} \delta^2}{(1+\delta^2)^2 t^2} &~\text{for}~\delta \neq 0
\end{cases}.
\end{split}
\ee
where we implicitly assume that $t \gg |X-x|$.
Thus, in the spatially non-compact system, in the large time regime of the non-unitary time evolution, i.e., $\delta t \gg 1$, those energy densities decay in time as $t^{-2}$, while for that of the unitary time evolution, i.e., $t\gg1$ with $\delta =0$, they decay as $t^{-4}$.
Unlike the chiral and anti-chiral energy densities in the compact system, in the non-compact system, the leading behaviors when $\delta t \gg 1$ depend on $\delta$.

Finally, let us interpret the late-time behaviors of the chiral and anti-chiral energy densities in (\ref{eq:late-EDS-inf}) with the large-time expansion of the density operator in (\ref{eq:large-time-expansion-of-rdo}).
Based on the results in the compact system, we further assume that $\rho^{(0)}_{X}(t)$ is independent of time.
Thus, to leading order in the large time expansion, the expectation values of the chiral and anti-chiral energy densities must vanish, i.e,
\be \label{eq:late-time-ED-inf}
\big\langle T (w_{X})\big\rangle_0=\Tr \left(\rho^{(0)} T (w_{X})\right)=0, \big\langle \overline{T} (\overline{w}_{X})\big\rangle_0=\Tr \left(\rho^{(0)} \overline{T} (\overline{w}_{X})\right)=0.
\ee
These late-time behaviors in (\ref{eq:late-time-ED-inf}) suggest that the late-time system is approximately the vacuum state with a small excitation corresponding to $\delta \rho (t)$\footnote{In the non-compact system, the energy density of the primary state collapses that of the vacuum one.
Therefore, the energy density may not distinguish the vacuum state from the primary one.  }.
Furthermore, this small excitation induces the late-time behavior of the chiral and anti-chiral energy densities reported in (\ref{eq:late-EDS-inf}).
\subsection{Second Rényi entanglement entropy \label{sec:SREE}}
In this section, we investigate whether the subsystems, and not just single spatial points, evolve to those of the primary state during the complex time evolution under consideration.

We now report the complex time dependence of the second Rényi entanglement entropy for the single interval with $A=[X_2,X_1]$.
Here, we consider the spatially-compact system with the circumference of $L$.
To quantify the amount of entanglement entropy induced by the complex time-evolved local operator, we define the change of the $n$-th R\'enyi entanglement entropy as
\be \label{eq:growth-of-REE}
\Delta S^{(n)}_{A}=S^{(n)}_A-S^{(n)}_{A;vac},
\ee
where $S^{(n)}_{A;vac}$ is defined as the $n$-th R\'enyi entropy for the vacuum state.
It should be pointed out that we will obtain the time dependence of (\ref{eq:growth-of-REE}) by analytically calculating the Euclidean counterpart of (\ref{eq:growth-of-REE}), which is denoted by $\Delta S^{(n)}_{A,E}$,  in the Euclidean path-integral formalism, and then performing an analytic continuation as $\tau_1=t$ and $\tau_2=it$.

Here, we consider the system described by $c=1$ free boson as a simple example.
Furthermore, as the inserted local operator, we choose the holomorphic current,
\begin{equation}
	\mathcal O(w,\overline{w})= i\,\partial_{w}\phi(w),\qquad (h_{\mathcal O},\overline{h}_{\mathcal O})=(1,0).
\end{equation}
Setting $n=2$, mapping from the cylinder to the plane where $\Delta S^{(2)}_A$ is expressed as the function of the standard cross ratios $(\eta_2, \overline{\eta}_2)$ (for details, see App.~\ref{app:uniformization}), and performing the analytic continuation as $\tau_1=t$, $\tau_2=it$,  in general systems described by the $2$d CFTs, we obtain the time dependence of $\Delta S^{(2)}_{A}$ as
\begin{equation}
	\label{eq:DeltaS2-master}
	\Delta S_A^{(2)}(t)
	=-\log\!\left\{\big[\eta_2(1-\eta_2)\big]^{\,2h_{\mathcal O}}\;
	\big[\overline{\eta}_2(1-\overline{\eta}_2)\big]^{\,2\overline{h}_{\mathcal O}}\;
	G(\eta_2,\overline{\eta}_2)\right\},
\end{equation}
where $G(\eta_2,\overline{\eta}_2)$ is the reduced four-point function (a weighted sum of conformal blocks, not a single block) determined by the CFT data.
The explicit forms, before and after the analytic continuation, of $(\eta_2,\overline{\eta}_2)$ are presented in App.~\ref{app:eta}. 
We divide $\Delta S^{(2)}_A$ into the theory-independent piece, $\Delta S^{(2)}_{A; \text{ind.}}$, and the theory-dependent piece, $\Delta S^{(2)}_{A; \text{dep.}}$, which are defined as
\be
\begin{split}
    &\Delta S^{(2)}_{A; \text{ind.}}=-\log{\left[\big[\eta_2(1-\eta_2)\big]^{2h_{\mathcal O}}
	\big[\overline{\eta}_2(1-\overline{\eta}_2)\big]^{2\overline{h}_{\mathcal O}}\right]},\\
    &\Delta S^{(2)}_{A; \text{dep.}}=-\log{\left[
	G(\eta_2,\overline{\eta}_2)\right]}.
\end{split}
\ee
For $\mathcal{O}(w,\overline{w})=i \partial_w \phi(w)$, this is simplified as 
\be
\begin{split}
    &\Delta S^{(2)}_{A; \text{ind.}}=-\log{\left[\big[\eta_2(1-\eta_2)\big]^{2}\right]},\\ 
&\Delta S^{(2)}_{A; \text{dep.}}
=-\log{\left[
	G(\eta_2)\right]}
=-\log{\left[
\zeta_{13}^{2}\zeta_{24}^{2}\,
\langle \mathcal O(\zeta_1)\mathcal O(\zeta_2)\mathcal O(\zeta_3)\mathcal O(\zeta_4)\rangle
\right]},
\end{split}
\ee
where $\zeta_{1,2,3,4}$ denote the images of the operator insertion points $w_i$ on the two-sheeted replica surface $\Sigma_2$ under the  map to the  $\zeta$-plane (for details, see Eq.~\eqref{eq:app-zeta}). 

We will report the time dependence of $\Delta S^{(2)}_A$ during the unitary time evolution.
As reported in \cite{Nozaki_2014,Nozaki:2014uaa,2014_Song}, during the unitary time evolution, $\Delta S^{(2)}_A$ is zero.

 Next, we move on to
 the late time behavior of $\Delta S^{(2)}_A$ for the free boson system with the insertion of $i\partial_z \phi (z)$ during the complex time evolution, i.e., $\delta \neq 0$. 
We start with $\Delta S^{(2)}_{A; \text{ind.}}$. 
If we expand the cross ratios with respect to   $\delta t\gg  1$ , the leading term 
is given by a function that is independent of time,
\begin{equation}
	\label{eq:eta-late}
	\eta_2\;\approx\;\sin^2\!\left[\frac{\pi(X_1-X_2)}{2L}\right],\qquad
1-\eta_2\;\approx\;1-\sin^2\!\left[\frac{\pi(X_1-X_2)}{2L}\right],
\end{equation}
Consequently, we obtain the late-time behavior of $\Delta S^{(2)}_{A; \text{ind.}}$ as 
\be \label{eq:2ndREE-ind}
\Delta S^{(2)}_{A; \text{ind.}}(t)
= -\log\!\left[\eta_2^2(1-\eta_2)^2\right]
 \underset{\delta t \gg 1}{\approx} 
 -2\log\!\left[\frac14\sin^2\!\left(\frac{\pi(X_1-X_2)}{L}\right)\right].
\ee
Next, we report the late-time behavior of $\Delta S^{(2)}_{A; \text{dep.}}$.
In the $c=1$ free boson system, by applying Wick's contraction, we obtain the holomorphic four-point function of $\mathcal O=i\partial\phi$ as
\begin{equation}
	\langle \mathcal O(\zeta_1)\mathcal O(\zeta_2)\mathcal O(\zeta_3)\mathcal O(\zeta_4)\rangle
	=\frac{1}{\zeta_{12}^2 \zeta_{34}^2}+\frac{1}{\zeta_{13}^2 \zeta_{24}^2}+\frac{1}{\zeta_{14}^2 \zeta_{23}^2},
\end{equation}
where $\zeta_{ij}=\zeta_i-\zeta_j$.
By using the cross ratio, $\eta_2$, we can express this four point function as
\begin{equation}
\begin{split}
  & \zeta_{13}^{2}\zeta_{24}^{2}\, \langle \mathcal O(\zeta_1)\mathcal O(\zeta_2)\mathcal O(\zeta_3)\mathcal O(\zeta_4)\rangle=G(\eta_2)
	=\frac{1}{\eta_2^2}+1+\frac{1}{(1-\eta_2)^2}
	=\frac{(1-\eta_2+\eta_2^2)^2}{\eta_2^2(1-\eta_2)^2}.\\
\end{split}
\end{equation}
Consequently, we obtain $\Delta S^{(2)}_{A; \text{dep.}}$ as 
\be
\begin{split}
    \Delta S^{(2)}_{A; \text{dep.}} =-\log{\left[\frac{(1-\eta_2+\eta_2^2)^2}{\eta_2^2(1-\eta_2)^2}\right]} \underset{\delta t \gg 1}{\approx}-2\log\!\left[\frac{1-\frac14\sin^2\!\left(\frac{\pi(X_1-X_2)}{L}\right)}{\frac14\sin^2\!\left(\frac{\pi(X_1-X_2)}{L}\right)}\right] .
\end{split}
\ee
where in the last equation, we take the leading term in the expansion with respect to $\delta t \gg1$.

By combining $\Delta S^{(2)}_{A;\text{ind.}}$ with $\Delta S^{(2)}_{A;\text{dep.}}$, we obtain $\Delta S^{(2)}_A$ as
\begin{equation}\label{eq:2nd-renyi-entropy}
	\Delta S_A^{(2)}(t)
	=-\log\!\left\{\left(1-\eta_2+\eta_2^2\right)^2\right\}
	=-2\log\!\left(1-\eta_2+\eta_2^2\right).
\end{equation}
By using the approximation in (\ref{eq:eta-late}), we obtain the late-time behavior of $\Delta S^{(2)}_{A}$ as
\begin{equation}\label{eq:final}
	\Delta S_A^{(2)}(t)\;\stackrel{\delta t \gg L}{\approx}\;
	-2\log\!\left[1-\sin^2\!\frac{\pi(X_1-X_2)}{2L}+\sin^4\!\frac{\pi(X_1-X_2)}{2L}\right]
	=-2\log\!\left[1-\frac14\sin^2\!\frac{\pi(X_1-X_2)}{L}\right].
\end{equation}
The pioneering papers \cite{Alcaraz:2011tn,2012JSMTE..01..016I} reported the entanglement entropy for the primary state corresponding to $i\partial_z \phi(z)$.
The late time behavior of $\Delta S^{(2)}_A$ in (\ref{eq:final}) matches  the findings in \cite{Alcaraz:2011tn,2012JSMTE..01..016I}.
This match suggests that 
the vacuum state with the insertion of $i\partial_w \phi(w)$ evolves to the primary state with the same conformal dimensions as those of  $i\partial_z \phi(z)$ 
under complex time evolution.
We also investigate the late-time value of $\Delta S^{(2)}_A$ for $\mathcal{O}=e^{\f{i}{2}\phi}$ and $\f{1}{\sqrt{2}}\left(e^{\f{i}{2}\phi}+e^{\f{-i}{2}\phi}\right)$, and find that it matches the primary-state entanglement entropy reported in\cite{Alcaraz:2011tn,2012JSMTE..01..016I}. 
This 
supports the inference that the complex time evolution evolves the vacuum state with 
a primary operator insertion
to the primary state system with the same conformal dimensions as the inserted operator. 
We postpone our findings for $\mathcal{O}=e^{\f{i}{2}\phi}$ and $\f{1}{\sqrt{2}}\left(e^{\f{i}{2}\phi}+e^{\f{-i}{2}\phi}\right)$ to App. \ref{app:O1O2}.



\begin{figure}[htbp]
	\centering
	\subfloat[$\delta = 0.05.$]{\includegraphics[width=.3\columnwidth]{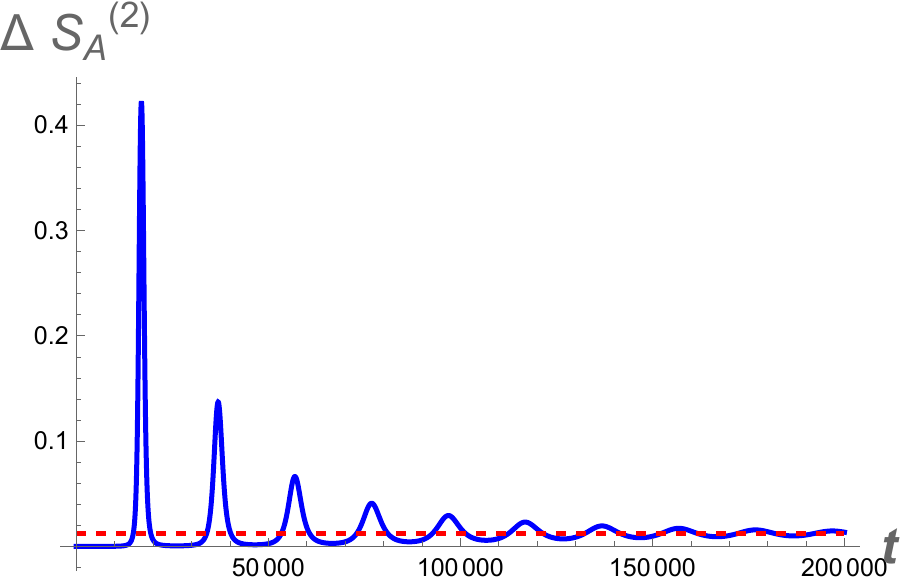}}\hspace{5pt}
	\subfloat[$\delta = 0.1.$]{\includegraphics[width=.3\columnwidth]{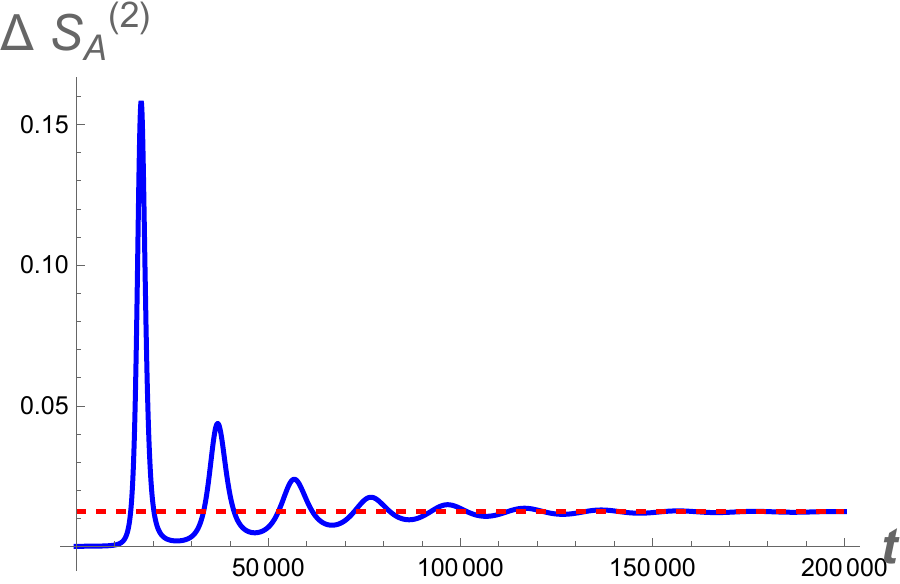}}\hspace{5pt}
	\subfloat[$\delta = 1.$]{\includegraphics[width=.3\columnwidth]{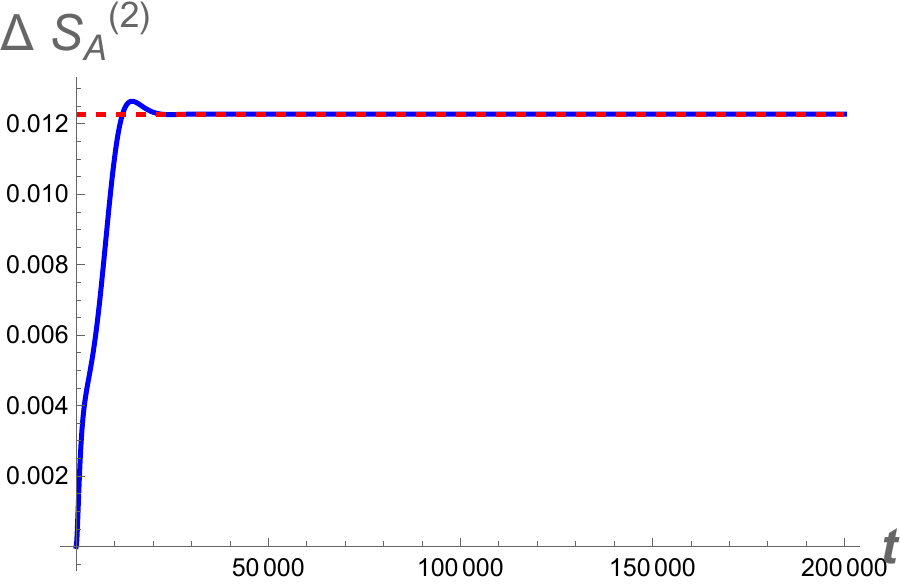}}\hspace{5pt}
	\caption{
		The growth of the second R\"enyi entanglement entropy, $\Delta S_A^{(2)}(t)$, for various $\delta$ as the function of $t$.
		In panels, (a)–(c), we show the complex time dependence of $\Delta S_A^{(2)}(t)$ for several values of $\delta$. 
        The parameters are set to be $\beta=10$ , $x=0$ 
		$A=[X_2,X_1]$, $X_2=2500$, $X_1=3500$, $L=20000$, and $c=1000$. 
        Here, the blue curve is the complex time dependence of $\Delta S^{(2)}_A$, while red curve is (\ref{eq:final}).  
        }
	\label{Fig:boson-1}
\end{figure}

We close this section with plots of $\Delta S^{(2)}_A$.
In Fig.~\ref{Fig:boson-1}, we show $\Delta S_A^{(2)}(t)$ for several $\delta$ as a function of $t$.
For $\delta =0.05$ and $0.1$, we observe the periodic behavior in $t$ of $\Delta S^{(2)}_A$.
Furthermore, for all $\delta$ under consideration, $\Delta S^{(2)}_A$ saturates to a certain value which is consistent with (\ref{eq:final}).

\subsection{Two-dimensional holographic entanglement entropy in the compact system \label{sec:2d-com-hol}}
In this section, we will further investigate if the complex time evolution, in the compact system, evolves the system from the vacuum state with the insertion of a local operator to the primary state with the same conformal dimensions as that of the inserted operator by studying the entanglement entropy, a quantity that is more commonly used to diagnose bipartite entanglement than the R\'enyi entanglement entropy.  
From this point on, we take the inserted operator to be a spinless one, i.e., $(h_{\mathcal O},\overline{h}_{\mathcal O})=(h_{\mathcal O},h_{\mathcal O})$.
Also, we consider a system described by two-dimensional holographic CFTs, the $2$d CFTs that admit gravitational duals, since in these system, we can easily calculate the entanglement entropy.
We may also investigate how the non-equilibrium phenomenon under consideration depends on the details of the CFTs by comparing these findings with Section \ref{sec:SREE}.

In $2$d holographic CFTs, we can obtain the entanglement entropy as the length of the geodesic anchored on the subsystem under consideration \cite{Ryu_2006,RT2006b}.
To perform the analytic calculation, we start from the three-dimensional Euclidean geometry dual to the Euclidean counterpart of our system,  calculate the length of the geodesics anchored on the boundary interval under consideration, and then perform the analytic continuation $\tau_1=t$, $\tau_2=it$ to obtain the
Lorentzian-time dependence of the entropy.

We begin with the
Euclidean gravity dual of (\ref{eq:euc-den-for-los}).
We assume that the Euclidean dual geometry is a vacuum solution of three-dimensional Einstein gravity with a negative cosmological constant. As shown in
\cite{Ban_ados_1999}, the most general solution with asymptotically AdS$_3$ boundary
conditions can be written as
\begin{equation}
	\label{eq:metric-banados}
	d s^2=\frac{L(w)}{2}\, d w^2
	+\frac{\overline{L}(\overline{w})}{2}\, d \overline{w}^2
	+\left(\frac{1}{\nu^2}
	+\frac{\nu^2}{4}\,L(w)\overline{L}(\overline{w})\right) d w\, d \overline{w}
	+\frac{d \nu^2}{\nu^2},
\end{equation}
where $\nu$ is the radial coordinate (with the conformal boundary at $\nu\to 0$), and
$(w,\overline{w})$ are the complex coordinates on the boundary cylinder introduced in the
previous section. The holomorphic and anti-holomorphic functions, $L(w)$ and
$\overline{L}(\overline{w})$,  are related to the chiral and anti-chiral components of the CFT stress tensor via
\begin{equation}
	\label{eq:LandT}
	L(w)=\frac{12\,\langle T(w)\rangle}{c},\qquad
	\overline{L}(\overline{w})=\frac{12\,\langle\overline{T}(\overline{w})\rangle}{c},
\end{equation}
where $c$ is the central charge and  $\left \langle T(w)\right \rangle $ and $\left \langle \overline{T}(\overline{w})\right \rangle$ are the chiral and anti-chiral energy densities in the $(w,\overline{w})$ coordinates respectively which are reported in (\ref{eq:ED-general}).
\if[0]
For the local operator quench considered in Sec.~\ref{sec:Energy-density}, these one-point
functions are given on the cylinder by
\begin{equation}
	\label{eq:ED-w}
	\begin{aligned}
		\big\langle T (w)\big\rangle 
		&=-\frac{c}{24}\!\left(\frac{2\pi}{L}\right)^2
		+ h_{\mathcal O}\!\left(\frac{dz}{dw}\right)^{\!2}
		\left[\frac{1}{z-z_1}-\frac{1}{z-z_2}\right]^{\!2},\\
		\big\langle \overline{T} (\overline{w})\big\rangle 
		&=-\frac{c}{24}\!\left(\frac{2\pi}{L}\right)^2
		+ h_{\mathcal O}\!\left(\frac{d\overline{z}}{d\overline{w}}\right)^{\!2}
		\left[\frac{1}{\overline{z}-\overline{z}_1}-\frac{1}{\overline{z}-\overline{z}_2}\right]^{\!2},
	\end{aligned}
\end{equation}
where $z=e^{\frac{2\pi}{L}w}$, $\overline{z}=e^{\frac{2\pi}{L}\overline{w}}$ and $(z_{1,2},\overline{z}_{1,2})$ 
denote the images of the local operator insertions on the plane. \fi

As reviewed in \cite{1986CMaPh.104..207B}, any Bañados geometry can be locally mapped to pure three-dimensional anti de Sitter space (AdS$_3$). 
The map from the Bañados geometry in $(w,\overline{w},\nu)$
to Poincaré AdS$_3$ in $(\tilde{w},\overline{\tilde{w}},\tilde{\nu})$ is
\begin{equation}
	\label{TS:BNA-ADS}
	\begin{split}
		\tilde{w} &= f(w)
		-\frac{2\nu^2 (f'(w))^2 \overline{f}''(\overline{w})}
		{4 f'(w) \overline{f}'(\overline{w})+\nu^2 f''(w) \overline{f}''(\overline{w})},\qquad
		\overline{\tilde{w}} = \overline{f}(\overline{w})
		-\frac{2\nu^2 (\overline{f}'(\overline{w}))^2 f''(w)}
		{4 f'(w) \overline{f}'(\overline{w})+\nu^2 f'' (w)\overline{f}''(\overline{w})},\\[4pt]
		\tilde{\nu} &= \frac{4\nu (f'(w)\overline{f}'(\overline{w}))^{3/2}}
		{4 f'(w) \overline{f}'(\overline{w})+\nu^2 f''(w) \overline{f}''(\overline{w})},
	\end{split}
\end{equation}
where $f(w)$ and $\overline{f}(\overline{w})$ are holomorphic and anti-holomorphic functions, and $f^{\overset{n}{\overbrace{'\cdots'}}}(w)=\f{d^nf(w)}{dw^n}$ and $\overline{f}^{\overset{n}{\overbrace{'\cdots'}}}(\overline{w})=\f{d^n\overline{f}(\overline{w})}{d \overline{w}^n}$. 
It should be pointed out that $\tilde{w}$ and $\overline{\tilde{w}}$ near the boundary, i.e, $\tilde{\nu}=\nu \approx 0$, can be approximated as the holomorphic and anti-holomorphic functions on the boundary, respectively,
\be \label{eq:boundary-map}
\tilde{w}\underset{\tilde{\nu}\approx 0}{\approx} f(w),~ \overline{\tilde{w}}\underset{\tilde{\nu}\approx 0}{\approx} \overline{f}(\overline{w}).
\ee
These coordinates, $(\tilde{w},\overline{\tilde{w}},\tilde{\nu})$ can express the geometry as pure Euclidean AdS$_3$,
\begin{equation}
	\label{eq:pure-ads-3}
	d s^2=\frac{d \tilde{w}\, d \overline{\tilde{w}}+d \tilde{\nu}^2}{\tilde{\nu}^2}.
\end{equation}
By using the map in (\ref{eq:boundary-map}), we obtain the chiral and anti-chiral energy densities as the Schwarzian derivatives.
Correspondingly, we obtain the relation between $L(w)$ and $\overline{L} (\overline{w})$ and $f(w)$ and $\overline{f} (\overline{w})$ as
 \begin{equation}
	\label{eq:schwarzian}
	\begin{split}
		L(w)&=\{f(w),w\}
		=\frac{3(f''(w))^2-2f'(w)f'''(w)}{2(f'(w))^2},\\
		\overline{L}(\overline{w})&=\{\overline{f}(\overline{w}),\overline{w}\}
		=\frac{3(\overline{f}(\overline{w})'')^2-2\overline{f}(\overline{w})'\overline{f}'''(\overline{w})}{2\left(\overline{f}'(\overline{w})\right)^2}.
	\end{split}
\end{equation}
Since the geodesic length is invariant under bulk diffeomorphisms, we can compute the
holographic entanglement entropy as the geodesic length in the pure AdS$_3$.
For a single interval $A=[X_2,X_1]$ on the boundary cylinder, the geodesic length takes the universal form as (see e.g. \cite{Roberts:2012aq})
\begin{equation}
	S_A=\frac{c}{12}\log\!\left[
	\frac{\,|f(w_1)-f(w_2)|^2\,|\overline{f}(\overline{w}_1)-\overline{f}(\overline{w}_2)|^2}
	{|f'(w_1)f'(w_2)\overline{f}'(\overline{w}_1)\overline{f}'(\overline{w}_2)|\,\epsilon_1^2\epsilon_2^2}
	\right],
	\label{eq:GeneralGeodesicLength}
\end{equation}
where we assume that endpoints of the geodesic is $(w_1,\overline{w}_1, \epsilon_2)$ and $(w_2,\overline{w}_2, \epsilon_2)$.

Now, we explain how to calculate the time dependence of the entanglement entropy for the general $2$d holographic CFT systems.
We divide the holographic entanglement entropy in (\ref{eq:GeneralGeodesicLength}) into the holomorphic and anti-holomorphic pieces as
\be
S_A=S^{\text{hol.}}(w_1,w_2)+S^{\text{anti-hol.}}(\overline{w}_1,\overline{w}_2).
\ee
Subsequently, we perform the analytic continuation, and $S^{\text{hol.}}(w_1,w_2)$, the holomorphic piece, becomes independent of $S^{\text{anti-hol.}}(\overline{w}_1,\overline{w}_2)$,  the anti-holomorphic piece.
These analytically-continued functions can have singularities that causes these functions to be multi-valued.
We can consider the points where the functions become singular as the branch points of a Riemann surface, where the number of sheets of this surface is determined by the structure of the functions.
Furthermore, the value of the function can be determined by which sheet the complex variables are located.
Let $i$ label the points where $S^{\text{hol.}}(w_1,w_2)$ becomes singular, and let $a$ label the points where $S^{\text{anti-hol.}}(\overline{w}_1,\overline{w}_2)$ becomes singular.
Moreover, let $n_i$ denote which sheet of the Riemann surface, associated to the $i$-th singular point, the holomorphic complex variable is located on, and let $\overline{n}_a$ denote which sheet of the Riemann surface, associated to the $a$-th singular point, the anti-holomorphic complex variable is located on.
As the analytically continued geodesic length depends on both $n_i$ and $\overline{n}_a$, the candidates of the holographic entanglement entropy take the form depending on $n_i$ and $\overline{n}_a$ as 
\be
S^{n_1, \cdots, n_k,\overline{n}_1, \cdots, \overline{n}_m}_c=S^{\text{hol}}_{n_1, \cdots, n_k}(w_1, w_2)+S^{\text{anti-hol}}_{\overline{n}_1, \cdots, n\overline{n}_m}(\overline{w}_1, \overline{w}_2),
\ee
where we assume that $S^{\text{hol}}_{n_1, \cdots, n_k}(w_1, w_2)$ has $k$ singular points, while $S^{\text{anti-hol}}_{\overline{n}_1, \cdots, n\overline{n}_m}(\overline{w}_1, \overline{w}_2)$ has $m$ singular points.

Next, we present the method employed in this paper for calculating the time dependence of the entanglement entropy. 
Define $\{n_i\}$ and $\{\overline{n}_a\}$ as the sets of $n_i$ and $\overline{n}_a$, respectively. 
In this paper, we assume that the initial locations of the complex variables are determined so that $\{n_i\}$ and $\{\overline{n}_a\}$ satisfy the initial conditions.
Then, we assume that the holographic entanglement entropy is continuous with respect to $t$.
Under these two assumptions, we take the minimal one from the set of the geodesic length as
\be
S_A(t)=\text{min}_{\{n_i\}, \{\overline{n}_a\}\in\mathbb Z}~S^{\{n_i\}, \{\overline{n}_a\}}_{A} (t),
\ee
where $\text{min}_{\{n_i\}, \{\overline{n}_a\}}$ means that we take the sets of $n_i$ and $n_a$ such that they follows the two conditions described above.

\if[0]
we perform the analytic continuation as $\tau_1=t$ and $\tau_2=it$.
\textcolor{blue}{After that, (\ref{eq:GeneralGeodesicLength}) can become the multi-valued function since (\ref{eq:GeneralGeodesicLength}) have singular points.}
The complex time evolution merely induces the time dependence of the endpoints as $(w_i(t),\overline{w}_i(t))$.

\textcolor{red}{Because $f$ and $\overline{f}$ are multi-valued functions on the spatial cylinder, each
boundary endpoint admits an infinite tower of sheets related by monodromies around the
spatial circle. At the initial time $t=0$, we choose for each endpoint a definite sheet
label $n_i\in\mathbb Z$ (with $i=1,2$), corresponding to going around the spatial circle
$n_i$ times; concretely, if $w_i(0)$ is the position of the endpoint at $t=0$, we may
identify
\begin{equation}
	f_{n_i}(w_i(0)) \equiv f\bigl(w_i(0)+2\pi i n_i\bigr),
\end{equation}
and similarly for $\overline{f}_{-n_i}(\overline{w}_i(0))$ in the anti-holomorphic sector. We then
 define $f_{n_i}(w_i(t))$ and $\overline{f}_{-n_i}(\overline{w}_i(t))$ for $t>0$ by analytic
continuation along the real-time trajectory of the endpoints. In other words,
$(n_1,n_2)$ are labels of the initial sheets at $t=0$, and for each choice of
$(n_1,n_2)$ we continuously track the corresponding images of the endpoints as time
evolves, even if the points move out of the original fundamental domain. Each such
choice of initial sheets $(n_1,n_2)$ therefore defines a distinct bulk geodesic
homologous to the interval $A$, with renormalized length
\begin{equation}
	S_A^{(n_1,n_2)}(t)=
	\frac{c}{12}\log\!\left[
	\frac{4\,|f_{n_1}(w_1(t))-f_{n_2}(w_2(t))|^2
		|\overline{f}_{-n_1}(\overline{w}_1(t))-\overline{f}_{-n_2}(\overline{w}_2(t))|^2}
	{|f'_{n_1}(w_1(t))f'_{n_2}(w_2(t))
		\overline{f}'_{-n_1}(\overline{w}_1(t))\overline{f}'_{-n_2}(\overline{w}_2(t))|}
	\right],
	\label{eq:branch-geodesic}
\end{equation}
where the anti-holomorphic branches are chosen at $t=0$ to be the complex conjugates of
the holomorphic ones and then continued smoothly in time.}

\textcolor{blue}{ After the analytic continuation, $\overline{f}(\overline{w})$ becomes independent of $f(w)$, i.e., $\overline{f} (\overline{w}) \neq \left(f(w)\right)^{*}$, where $(\cdot)^{*}$ is complex conjugate of $(\cdot)$.
If 
The multi-valuedness of $f$ and $\overline{f}$ emerges as the branch of the Riemann surface where the singular points can be the branch points
Let $n_i \in \mathbb Z$ labels the sheets of Riemann surface associated with .
Furthermore, we assume if the complex variables are on the $m_i$-th sheet ($m_i$-th sheet), then they are shifted by $i2\pi m_i$. 
Define the functions associated with the complex variable on the $m_i$-the sheet as  
\be
f_{m_i}(w_i(t)) \equiv f\bigl(w_i(t)+2\pi i m_i\bigr), \overline{f}_{m_i}(\overline{w}_i(t)) \equiv \overline{f}\bigl(\overline{w}_i(t)+2\pi i m_i\bigr),
\ee
where the behavior of $f_{m_i}(w_i(t))$ and $\overline{f}_{m_i}(\overline{w}_i(t))$ depends on $m_i$.
}

\textcolor{red}{At any fixed complex time $t$, the physical holographic entanglement entropy is obtained
by selecting the shortest such geodesic, in accordance with the Ryu--Takayanagi
prescription,
\begin{equation}
	S_A(t)=\min_{n_1,n_2\in\mathbb Z}\; S_A^{(n_1,n_2)}(t).
	\label{eq:min-geodesic}
\end{equation}
In practice, the monodromy exponents entering $f$ and $\overline{f}$ make geodesics with large
$|n_i|$ exponentially long, so only a small window of branch labels around the origin
contributes. In our numerical implementation, we first fix the branch labels $(n_1,n_2)$
at $t=0$ for the two endpoints, then track for each pair $(n_1,n_2)$ the corresponding
geodesic length $S_A^{(n_1,n_2)}(t)$ as a function of time by analytic continuation of
$f_{n_i}(w_i(t))$ and $\overline{f}_{-n_i}(\overline{w}_i(t))$. At each time $t$ we take the minimum
over this finite set of branches as in \eqref{eq:min-geodesic}, which realizes the RT
prescription for the time-dependent Bañados geometry. The complex time dependence of
$S_A(t)$ is thus obtained from the analytic continuation $\tau_1\to t$, $\tau_2\to it$ of
the boundary insertion points, which induces the corresponding evolution of
$(w_i(t),\overline{w}_i(t))$ in the above formulas.}
\fi

Now, we move on the system under consideration.
We begin with the presentation of the single-interval entanglement entropy for the primary state.
For primary state with equal chiral and anti-chiral conformal dimensions, $(h_{\mathcal{O}},h_{\mathcal{O}})$, the chiral and anti-chiral energy densities are obtained as (\ref{eq:ED-late}).
By substituting (\ref{eq:ED-late}) into (\ref{eq:metric-banados}) and calculating the geodesic length as we explained above, then we obtain the holographic entanglement entropy as 
\be \label{eq:late-EE-prim}
S_A=  \frac{c}{3} \log \left(\frac{L \sinh \left(\frac{\pi \sqrt{\frac{24 h_{\mathcal{O}}}{c}-1}\left|X_1-X_2\right|}{L}\right)}{\pi \sqrt{\frac{24 h_{\mathcal{O}}}{c}-1}}\right).
\ee
Then, let us investigate the time dependence of the single-interval entanglement entropy for a given subsystem. 
When we take $A=[X_2,X_1]$ as the subsystem 
under consideration, we obtain the time dependence of $S_A$ as 
{\footnotesize
\begin{equation}\label{eq:TE-EE-con}
\begin{aligned}
S_A(t)= & \frac{c}{12} \ln \left\{\frac{L^4}{16 \pi^4\left(-1+\frac{24 h}{c}\right)^2 \sinh ^4\left(\frac{2 \pi}{L}(\beta+t \delta)\right)}\right. \\
& \times\left[\cosh \left(\frac{2 \pi}{L}(\beta+t \delta)\right)-\cos \left(\frac{2 \pi\left(t-X_1\right)}{L}\right)\right]\left[\cosh \left(\frac{2 \pi}{L}(\beta+t \delta)\right)-\cos \left(\frac{2 \pi\left(t+X_1\right)}{L}\right)\right] \\
& \times\left[\cosh \left(\frac{2 \pi}{L}(\beta+t \delta)\right)-\cos \left(\frac{2 \pi\left(t-X_2\right)}{L}\right)\right]\left[\cosh \left(\frac{2 \pi}{L}(\beta+t \delta)\right)-\cos \left(\frac{2 \pi\left(t+X_2\right)}{L}\right)\right] \\
& \times\left(e^{2 \sqrt{-1+\frac{24 h}{c}} \arctan \left(\operatorname{coth} \frac{\pi(\beta+t \delta)}{L} \tan \frac{\pi\left(t-X_1\right)}{L}\right)}-e^{2 \sqrt{-1+\frac{24 h}{c}} \arctan \left(\operatorname{coth} \frac{\pi(\beta+t \delta)}{L} \tan \frac{\pi\left(t-X_2\right)}{L}\right)}\right)^2 \\
& \left.\times\left(e^{-2 \sqrt{-1+\frac{24 h}{c}} \arctan \left(\operatorname{coth} \frac{\pi(\beta+t \delta)}{L} \tan \frac{\pi\left(t+X_1\right)}{L}\right)}-e^{-2 \sqrt{-1+\frac{24 h}{c}} \arctan \left(\operatorname{coth} \frac{\pi(\beta+t \delta)}{L} \tan \frac{\pi\left(t+X_2\right)}{L}\right)}\right)^2\right\} .
\end{aligned}
\end{equation}}In the late limit, where we expand (\ref{eq:TE-EE-con}) with respect to $\delta t\gg1$, 
the leading order term gives
\begin{equation} \label{eq:TE-EE-con-late}
	\begin{aligned}
		S_A (t \rightarrow \infty)
		&= \frac{c}{3} \log \left(\frac{L \sinh \left(\frac{\pi \sqrt{\frac{24 h_{\mathcal{O}}}{c}-1}\left|X_1-X_2\right|}{L}\right)}{\pi \sqrt{\frac{24 h_{\mathcal{O}}}{c}-1}}\right),
	\end{aligned}
\end{equation}
where the first term depends on the primary state, while the second term does not, and it is the same as the single-interval entanglement entropy for the ground state.
Thus, the late-time behavior of the single-interval entanglement entropy is the same as (\ref{eq:late-EE-prim}).
Furthermore, we compare our result with the result reported by \cite{Alcaraz:2011tn}.
In \cite{Alcaraz:2011tn}, they investigated the entanglement entropy for the primary state in the small interval limit, i.e., $(X_1-X_2)/L \ll 1$. 
The authors assumed that operator expansion of the primary operator is approximated as the identity operator, i.e., $\mathcal{O}\times \mathcal{O}^{\dagger}={\bf 1} +\cdots$, in this small interval limit.
To leading order of this small interval expansion with this assumption, the entanglement entropy was obtained as 
\be
S_{A}\approx \f{4\pi^2}{3}h_{\mathcal{O}}\left(\f{X_1-X_2}{L}\right)^2+ \frac{c}{3} \log \left(\frac{L}{\pi } \sin{\left[\frac{\pi\left(X_1-X_2\right)}{L}\right]}\right).
\ee
This is consistent with (\ref{eq:TE-EE-con-late}) at the leading order in the small interval expansion.

As discussed in Section \ref{sec:Energy-density}, by using the late-time reduced density operator, we will interpret the late-time behavior of the entanglement entropy reported in (\ref{eq:TE-EE-con-late}).
 To this end, we expand the late-time density operator as in (\ref{eq:large-time-expansion-of-rdo}), and at leading order in this expansion, we obtain the late-time entanglement entropy as
 \be
S_{A}=-\Tr_A\left[ \rho^{(0)}_A \log{\rho^{(0)}_A}\right].
 \ee
Since this should be the same as (\ref{eq:TE-EE-con-late}), the leading term of $\rho_A$ in the late-time region should be the same as the reduced density matrix for the primary state with the same conformal dimensions as those of the inserted operator.
In other words, from (\ref{eq:TE-EE-con-late}), we can see that the complex time evolution, induced by the $2$d holographic CFT Hamiltonian, evolves the subsystems from those of the vacuum state with an insertion of a local operator to those of the primary state with the same conformal dimensions as the ones of the inserted operator as in the case of the $2$d free CFT.

\begin{figure}[htbp]
	\centering 
	\subfloat[$\delta =0 , X_2=3000, X_1= 6000.$]{\includegraphics[width=.45\columnwidth]{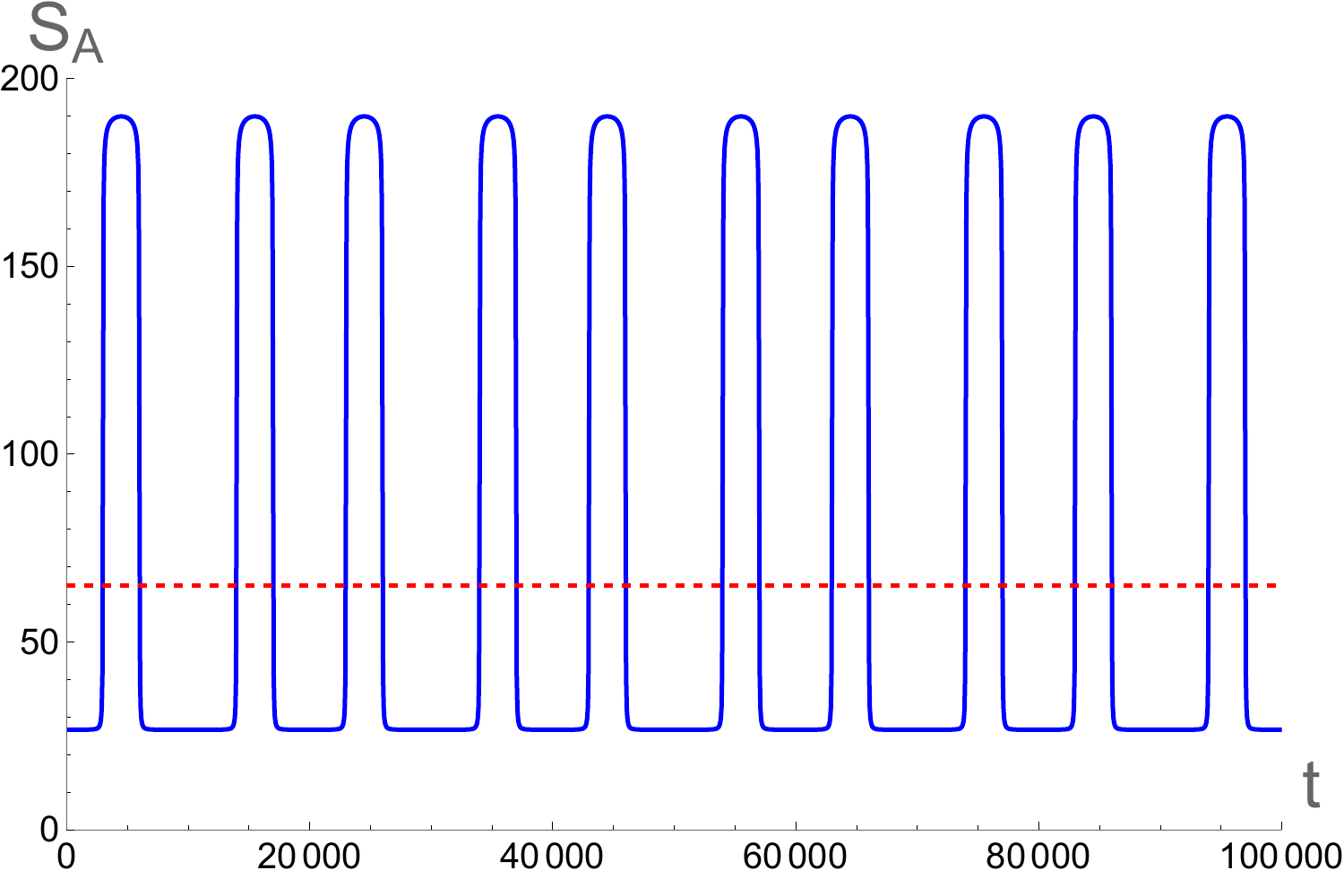}}\hspace{5pt}
	\subfloat[$\delta =0.01, X_2=3000, X_1= 6000.$]{\includegraphics[width=.45\columnwidth]{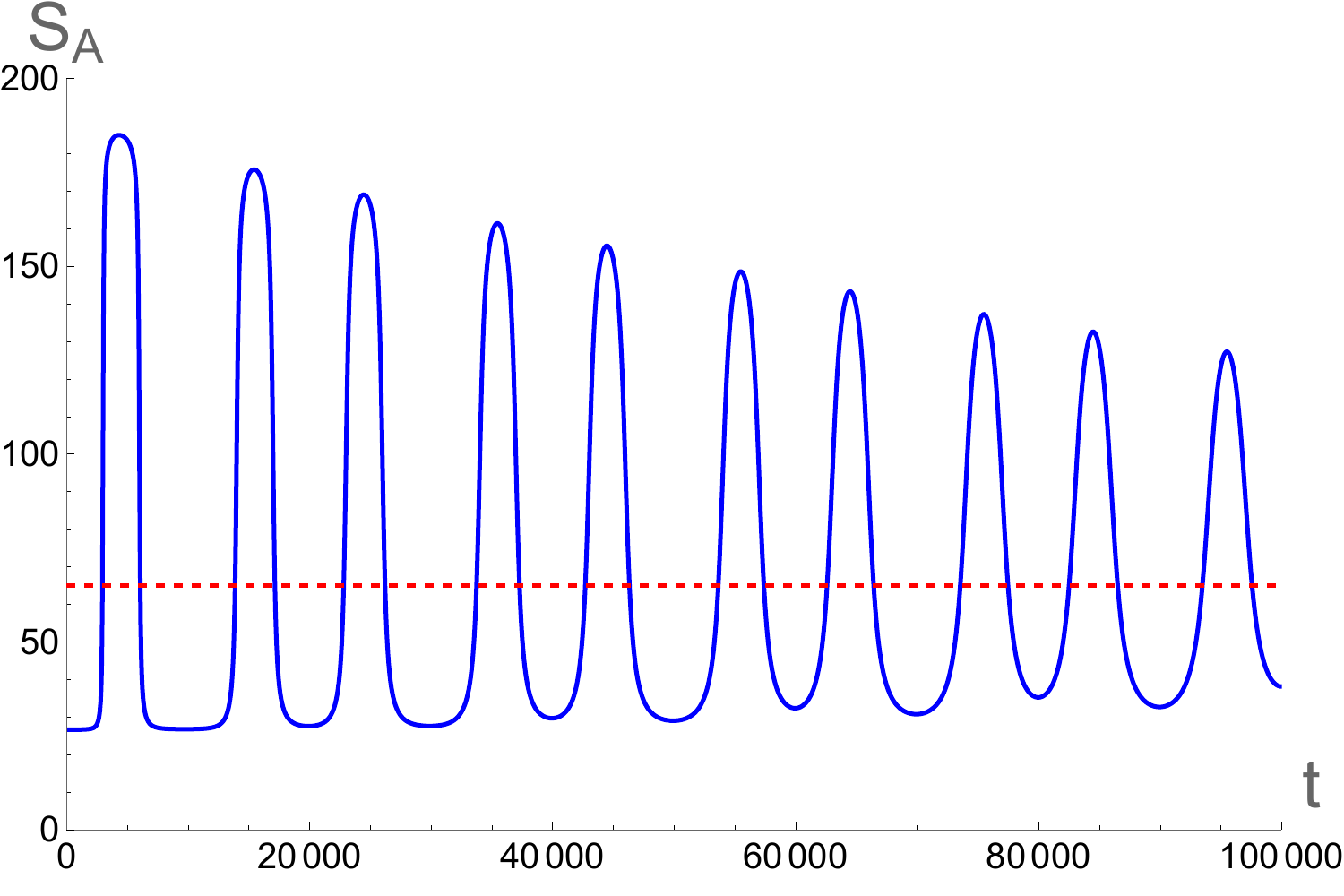}}\hspace{5pt}
	
	\subfloat[$\delta =0.1, X_2=3000, X_1= 6000.$]{\includegraphics[width=.45\columnwidth]{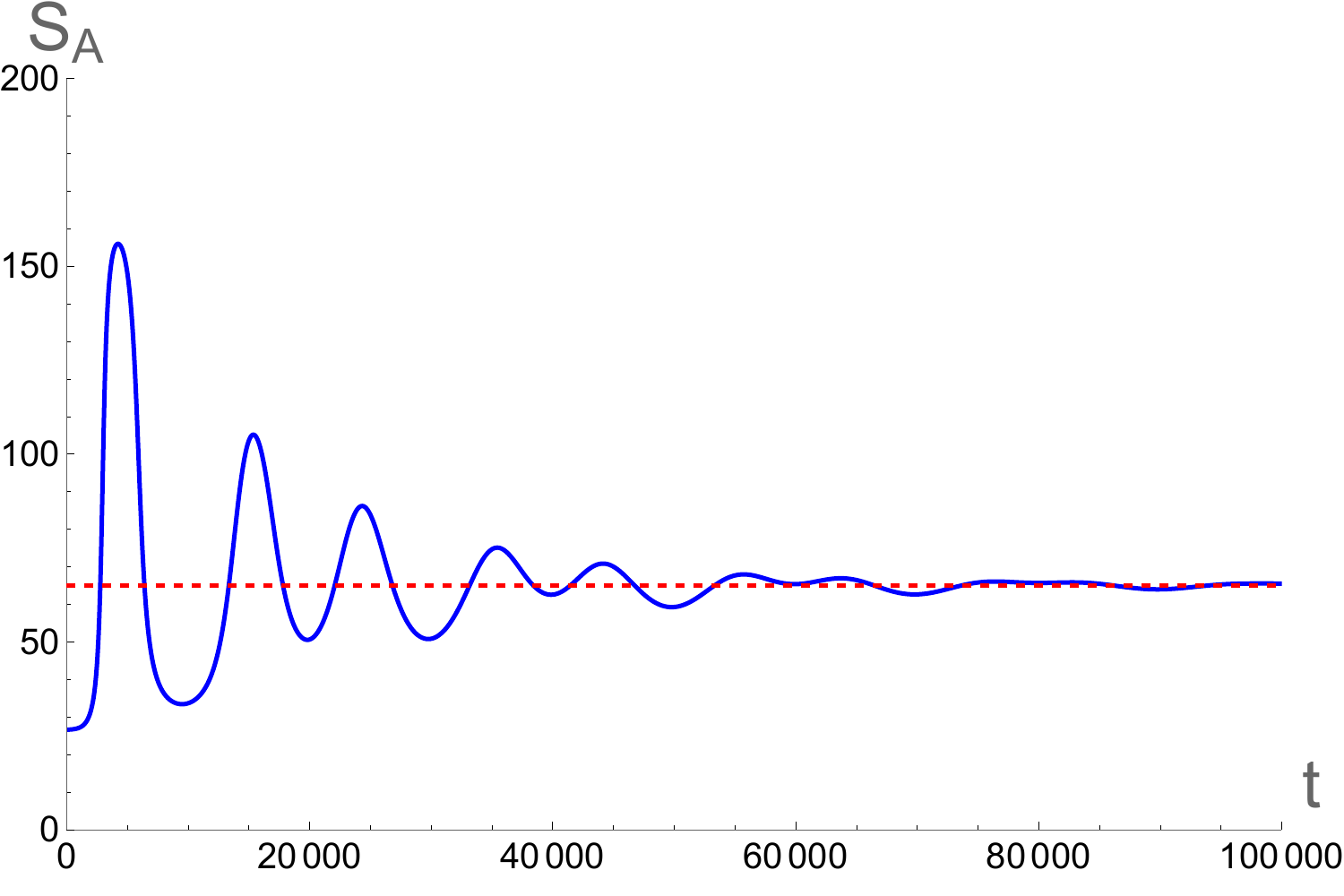}}\hspace{5pt}
	\subfloat[$\delta =1, X_2=3000, X_1= 6000.$]{\includegraphics[width=.45\columnwidth]{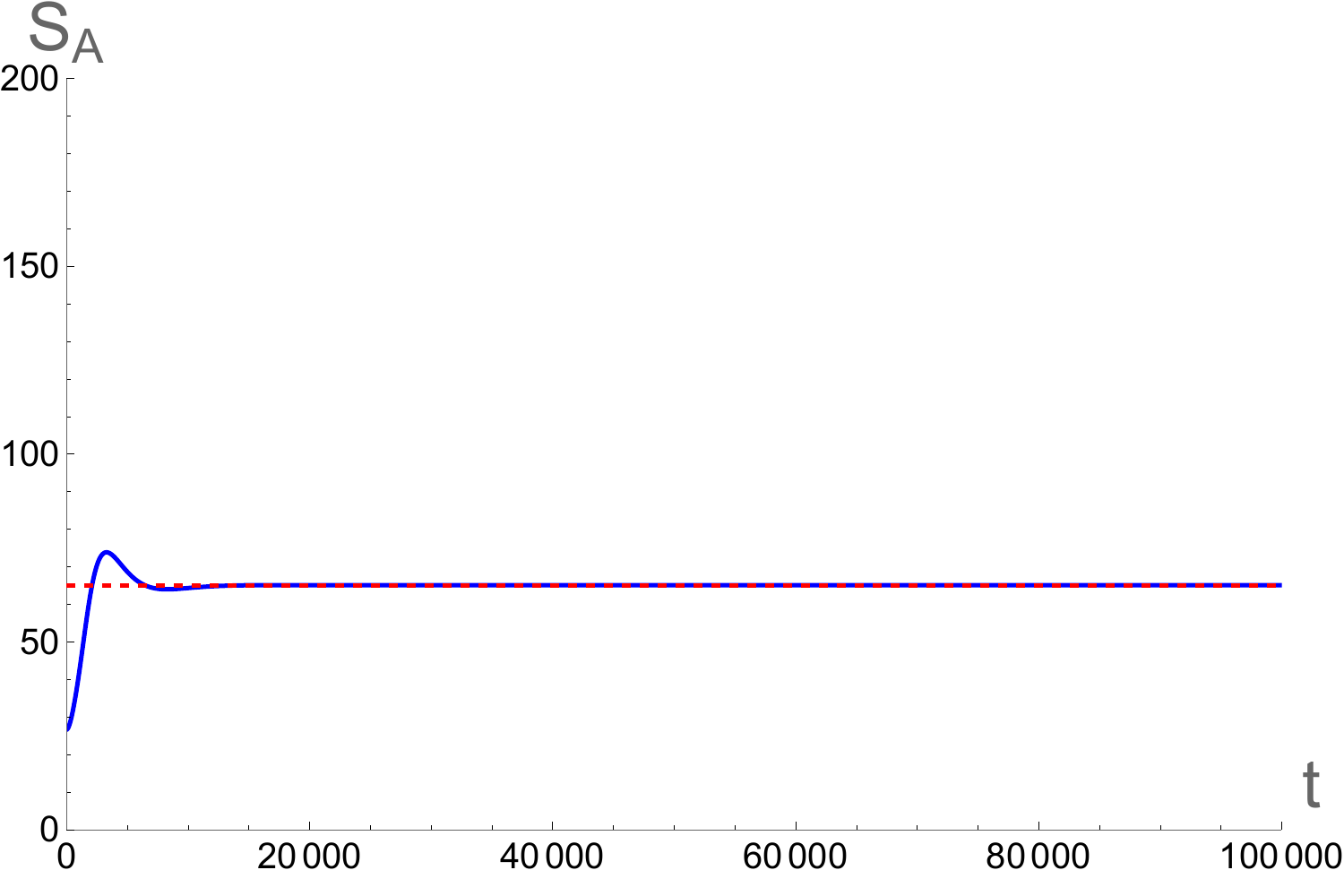}}
	\caption{ Entanglement entropy for various values of the deformation parameter~\(\delta\), as a function of $t$. Here, we set $\beta =20, x=0,  L=20000 , h_\mathcal{O}= \overline{h}_\mathcal{O} =1000, c=6$.
   The blue curve is the complex time dependence of $S_A$ and the red curve is \eqref{eq:late-EE-prim}.
	}\label{Fig:Geo-circle}
\end{figure}

Next, we report the time dependence of the single-interval entanglement entropy for several values of $\delta$.
In Fig. \ref{Fig:Geo-circle}, we show the time dependence of the entanglement entropy during the complex time evolution with several values of $\delta$.
From this figure, we can observe that during the unitary time evolution, i.e., $\delta =0$, the time dependence of the single-interval entanglement entropy exhibits the persistent periodic behavior with the period of $L$.
In contrast, we can observe that during the non-unitary time evolution, i.e., $\delta \neq 0$, the entanglement entropy oscillates and grows in time to that for the primary state with $(h_{\mathcal{O}}, h_{\mathcal{O}})$. 
As \(\delta\) increases, the oscillations damp out more rapidly and the entropy reaches its saturation value in a shorter time.

\subsubsection*{Formula Bridging Entanglement Entropy and Energy Densities}
Two of the authors in this paper reported the formula bridging the holographic entanglement entropy and the energy densities \cite{Mao:2025hkp}. 
By exploiting this formula, we can easily see why the holographic entanglement entropy saturates to the primary state with the same conformal dimensions as the inserted operator during the non-unitary time evolution.
This formula states that the holographic entanglement entropy is given as a function depending on only the chiral and anti-chiral energy densities.
As shown in (\ref{eq:ED-late}), the late-time behaviors of the chiral and anti-chiral energy densities are given by those of the primary state.
Therefore, as the late-time behavior of the holographic entanglement entropy is given by functions determined by only the chiral and anti-chiral energy densities for the primary state, the late-time holographic entanglement entropy is the entanglement entropy of this state.

\subsection{2d Holographic CFT on Infinite Space \label{sec:2dhCFT-inf}}
We conclude this section by looking at the time dependence of the entanglement entropy of a single interval in a spatially-infinite system described by $2$d holographic CFTs after the complex time quench. We will also report on our finding in the $2$d Ising CFT  in App \ref{sec:2d-ISing-CFT}.
As in the previous sections, we begin with the Euclidean calculation of the holographic entanglement entropy as geodesic length \eqref{eq:GeneralGeodesicLength}. 
Subsequently, by performing the analytic continuation $\tau_1\to t$, $\tau_2\to it$, we analytically investigate the complex time dependence of the entanglement entropy with the method of calculating it, which is reported in Section \ref{sec:2d-com-hol}.

First, we report the complex time dependence of the entanglement entropy for the finite interval, $A=[X_2,X_1]$.
Here, we expand the entanglement entropy in (\ref{eq:TE-EE-con}) with respect to $L \gg 1$, and take its leading term in this expansion.
Then, we treat this leading term as the complex time dependence of the entanglement entropy in the spatially-infinite system.

\begin{figure}[htbp]
	\centering
	\subfloat[$\delta =0$]{\includegraphics[width=.45\columnwidth]{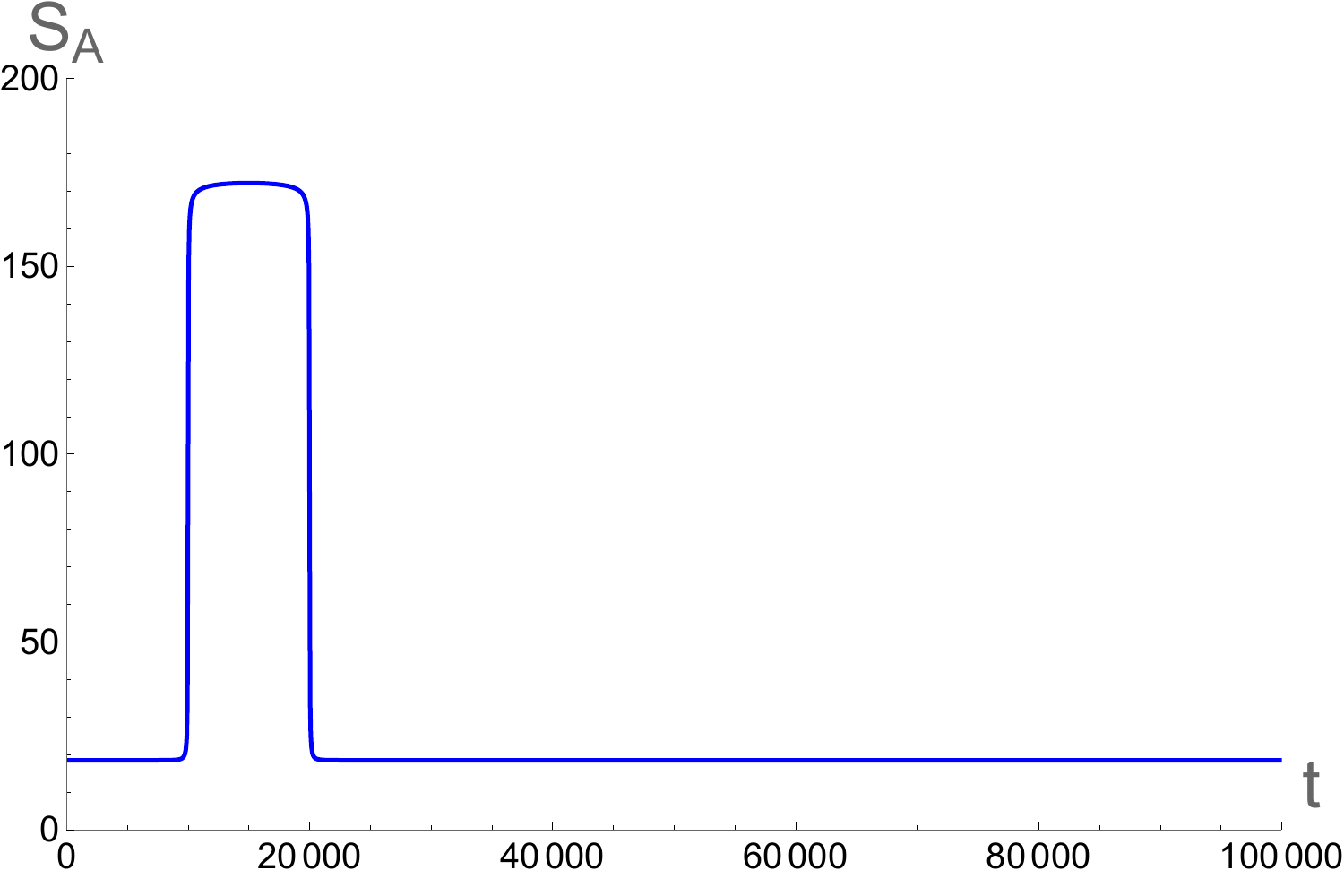}}\hspace{5pt}
	\subfloat[$\delta =0.01$]{\includegraphics[width=.45\columnwidth]{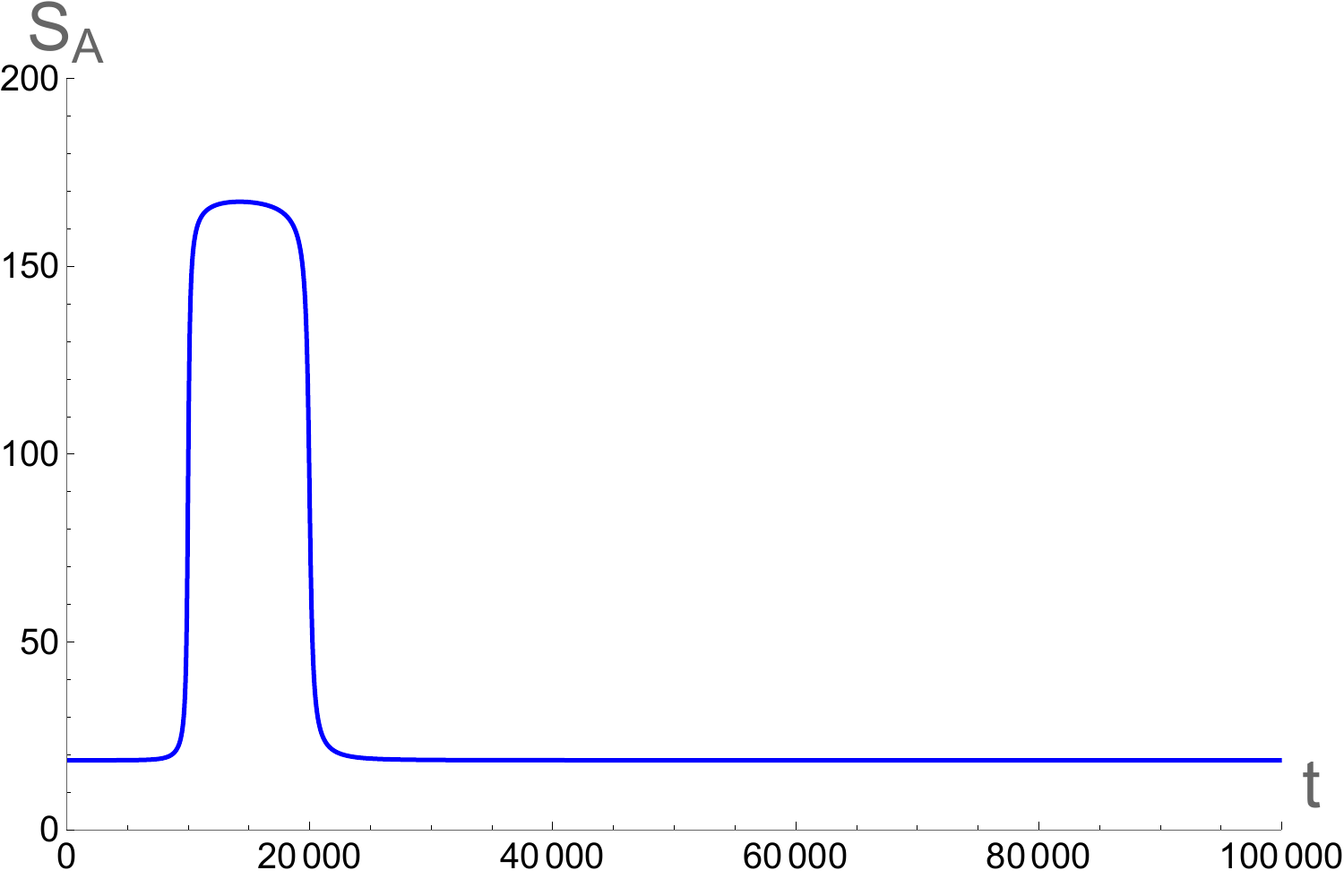}}\hspace{5pt}\\
	\subfloat[$\delta =0.1$]{\includegraphics[width=.45\columnwidth]{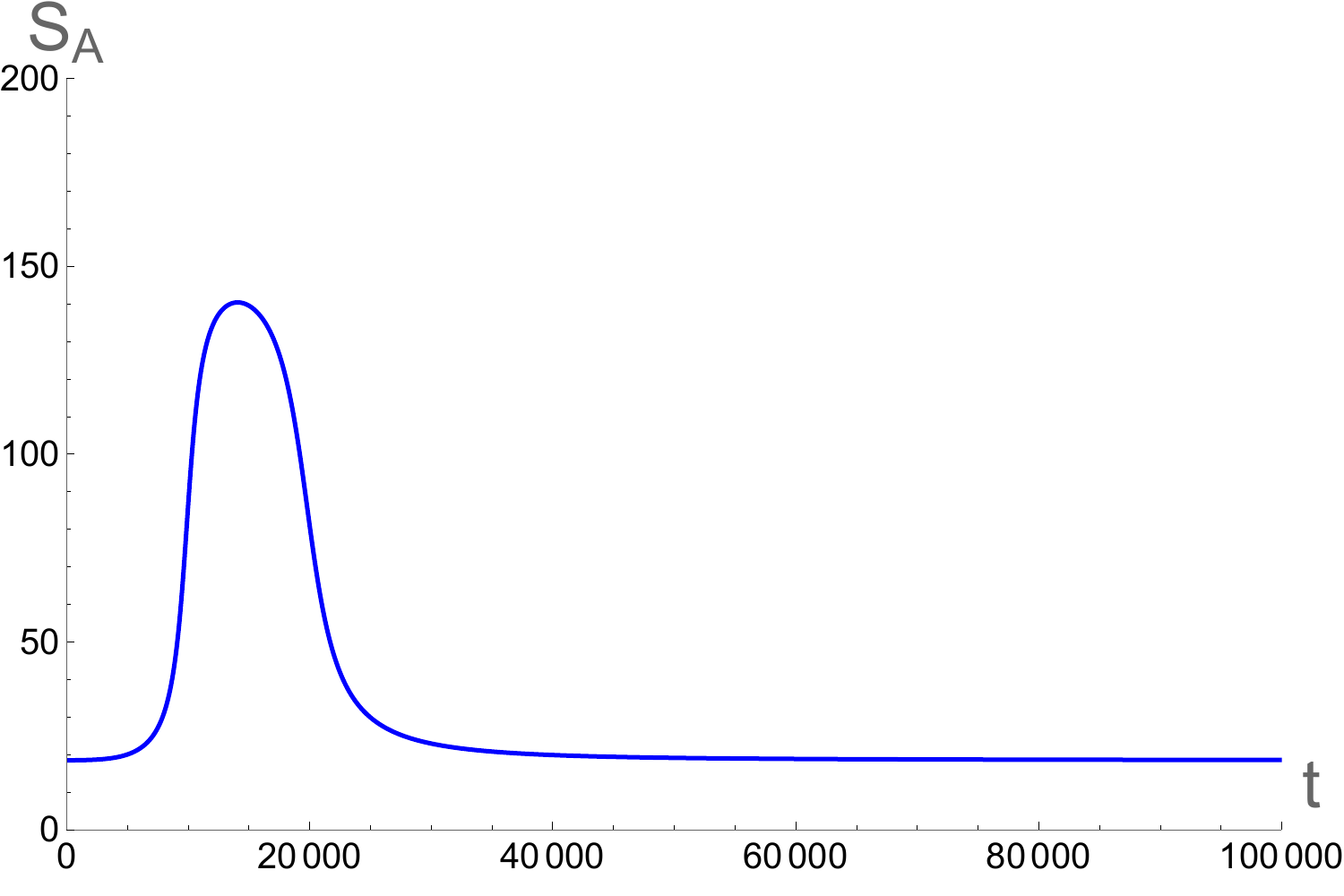}}\hspace{5pt}
	\subfloat[$\delta =1$]{\includegraphics[width=.45\columnwidth]{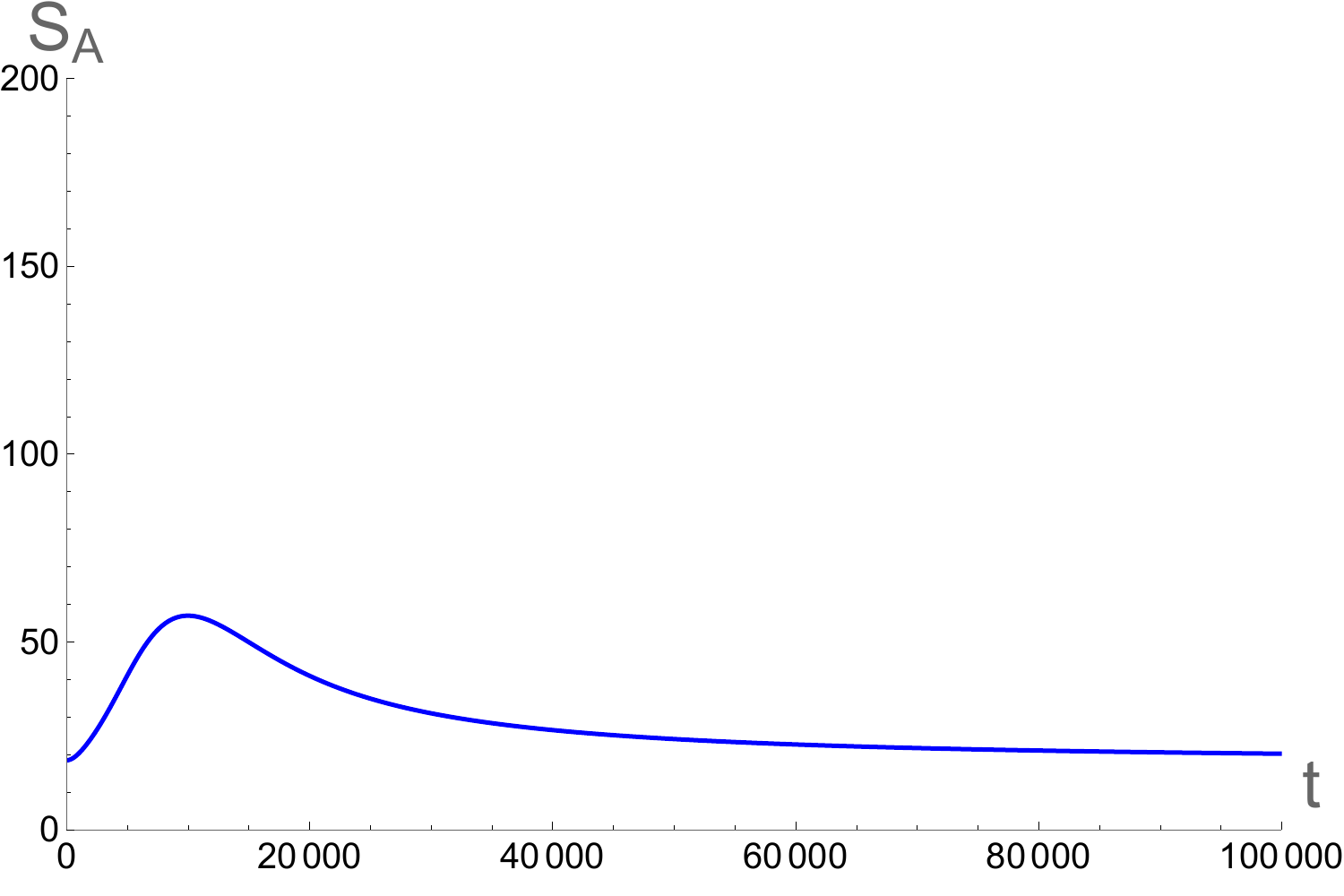}}
	\caption{Finite-interval entanglement entropy on infinite system with various $\delta$, as a function of $t$.
In this figure, we set $\beta =20, X_2=10000, X_1= 20000 ,h_\mathcal{O}= \overline{h}_\mathcal{O} =1000,$ and $c=1000$.
    }\label{Fig:Geo-Plane}
\end{figure}

In Fig. \ref{Fig:Geo-Plane}, we show the time dependence of $S_A$ for the various values of $\delta$.
In panel (a), we show the time dependence of the entanglement entropy during the complex time evolution with $\delta=0$, i.e., the unitary time evolution. 
From this figure, we can observe that only in the time interval, $X_2<t<X_1$, the entanglement entropy is larger than the vacuum entanglement entropy, while except for that time region, it is the vacuum one. 
In other words, the time dependence of the entanglement entropy exhibits a plateau in this time interval. 
To qualitatively describe the time dependence of the entanglement entropy during the unitary time evolution, we briefly introduce a quasiparticle picture \cite{Nozaki_2014,Nozaki:2014uaa}.
In this picture, we assume that the inserted local operator induces a pair of quasiparticles that are entangled with each other. 
One of them propagates to the left with the speed of light, while another propagates to the right with the same speed.
When the subsystem under consideration contains only a single quasiparticle, the entanglement between them can contribute to the entanglement entropy of the subsystem, and the value of the entanglement entropy becomes larger than the vacuum one.
This quasiparticle picture can describe the time dependence of $S_A$ in (a).

In panels (b)-(d), we show the time dependence of the entanglement entropy during the complex time evolution with $\delta >0$, i.e., the non-unitary time evolution.
One possible interpretation of this non-unitary time evolution is that during this evolution, we may replace $\beta$, the damping factor that tames the high-energy modes, with $\beta_{\text{eff}}=\beta+\delta t$, the time-dependent damping factor. 
The high-energy region that $\beta_{\text{eff}}$ tames may expand with time. 
From (b)-(d) in Fig. \ref{Fig:Geo-Plane}, we can observe that as $\delta$ increases, the plateau becomes less pronounced. 
The complex time dependence of $S_A$ with large $\delta$ exhibits a peak with a tail decaying in time, instead of the plateau.
This suggests that the entanglement dynamics induced by the complex time evolution with large $\delta$ cannot be effectively described by the propagation of the local objects, such as quasiparticles.
For the large times, the entanglement entropy saturates to the vacuum one.

\begin{figure}[htbp]
	\centering {\includegraphics[width=.45\columnwidth]{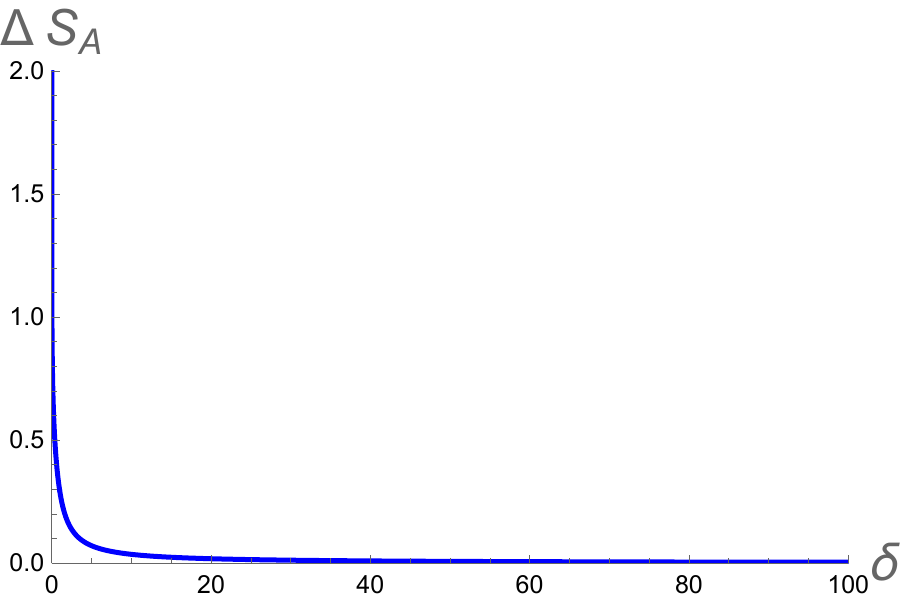}}\hspace{5pt}
	\caption{The $\delta$-dependence of the late-time value of $\Delta S_A$ for the semi-infinite interval. 
    Here, we set $c=1$ and $h_\mathcal{O}= \overline{h}_\mathcal{O} =1$. }
	\label{Fig:deltadenpendence}
\end{figure}

Next, we move on to the complex time evolution of the bipartite entanglement associated with the semi-infinite interval, $A=[X_2,\infty)$.
Let us define the growth of the entanglement entropy as
\be
\Delta S_A =S_A(t) -S^{\text{Vac}}_A,
\ee
where $S_A(t)$ is the entanglement entropy under consideration, while $S^{\text{Vac}}_A$ is the vacuum one.
Since the entanglement entropy for the semi-infinite interval can be infinite, we investigate the complex time evolution of $\Delta S_A$, instead of $S_A(t)$.
Here, we closely look at the late-time behavior of $\Delta S_A$ for the semi-infinite interval.
To this end, we begin with $\Delta S_A$ for $A=[X_2,X_1]$ in the spatially-infinite system.
Then, we expand it with respect to $X_1 \gg 1$, and take the leading term in this expansion as the complex time evolution of $\Delta S_A$ for the semi-infinite interval.
Furthermore, we expand it with respect to $t\gg 1$, and take the leading order in this expansion as the late-time behavior of the semi-interval $\Delta S_A$.
For $\delta=0$, the late-time $\Delta S_A$ logarithmically grows in time as 
\begin{equation}
	\Delta S_A
	\underset{t \gg 1}{\approx }\frac{c}{6}\,\log\!\left[
	\frac{t\,\sinh\!\Big(\pi\sqrt{h_{\mathcal{O}}-\tfrac{c}{24}}\Big)}
	{\beta\,\sqrt{h_{\mathcal{O}}-\tfrac{c}{24}}}
	\right],
\end{equation}
where the logarithmic growth is consistent with that reported in \cite{2014arXiv1405.5946C}.
For $\delta > 0$, the late–time value of $\Delta S_A$ saturates to a constant as

\begin{equation}\label{eq:DeltaS_late_sinh_ratio}
\Delta S_A\underset{t \gg 1}{\approx }\frac{c}{6}\,
\log\!\left[
\frac{
\sinh\!\Big(\sqrt{\,h_{\mathcal O}-\frac{c}{24}\,}\,\big(\pi-\arctan\delta\big)\Big)
}{
\sinh\!\Big(\sqrt{\,h_{\mathcal O}-\frac{c}{24}\,}\,\arctan\delta\Big)
}
\right].
\end{equation}
Differentiating \eqref{eq:DeltaS_late_sinh_ratio} with respect to $\delta$, 
we obtain
\begin{equation}
\label{eq:dDeltaS_ddelta_from_sinh}
\frac{d}{d\delta}\Delta S_A
\underset{t \gg 1}{\approx }
-\frac{c}{6}\,
\frac{\sqrt{\,h_{\mathcal O}-\frac{c}{24}\,}}{1+\delta^2}\,
\left[
\coth\!\Big(\sqrt{\,h_{\mathcal O}-\frac{c}{24}\,}\,\big(\pi-\arctan\delta\big)\Big)
+
\coth\!\Big(\sqrt{\,h_{\mathcal O}-\frac{c}{24}\,}\,\arctan\delta\Big)
\right].
\end{equation}
For $\delta>0$, one has $\arctan\delta\in(0,\pi/2)$, hence both arguments of $\coth$
in \eqref{eq:dDeltaS_ddelta_from_sinh} are strictly positive, and thus each $\coth$ term is positive.
Therefore, for $\delta>0$, $\Delta S_A$ decreases monotonically with $\delta$, i.e.,
\begin{equation}
\frac{d}{d\delta}\Delta S_A<0\qquad(\delta>0).
\end{equation}
Furthermore, in the limit $\delta\to +\infty$, $\Delta S_A$ reduces to zero as 
\begin{equation}
\lim_{\delta\to +\infty}\Delta S_A
=\frac{c}{6}\log\!\left[
\frac{\sinh\!\big(\sqrt{\,h_{\mathcal O}-\frac{c}{24}\,}\,\frac{\pi}{2}\big)}
{\sinh\!\big(\sqrt{\,h_{\mathcal O}-\frac{c}{24}\,}\,\frac{\pi}{2}\big)}
\right]=0.
\end{equation}
In Fig. \ref{Fig:deltadenpendence}, we show the $\delta$-dependence of $\Delta S_A$. 
From this figure, we can observe that $\Delta S_A$ monotonically decreases, according to the growth of $\delta$, so that it saturates to zero.


Thus, in contrast to the late-time behavior of $S_A$ in the compact system (see (\ref{eq:TE-EE-con-late})), that of $\Delta S_A$ in the spatially-infinite system depends on $\delta$. 

\subsubsection*{Formula Bridging Entanglement Entropy and Energy Densities}
By exploiting the formula bridging the entanglement entropy and the chiral and anti-chiral energy densities, which was reported in \cite{2025arXiv250807645M}, we discuss what induces the $\delta$-dependence of the entanglement entropy. 
As mentioned in the previous section, the formula proposed in \cite{2025arXiv250807645M} states that holographic entanglement entropy is a function determined by the chiral and anti-chiral energy densities.
First, we discuss what induces the system dependence of the late-time entanglement entropy with $\delta \neq 0$.
In the compact system, the chiral and anti-chiral energy densities saturate to those for the primary state with the same conformal dimensions as those of the inserted primary operator during the complex time evolution.
Correspondingly, the entanglement entropy for the finite interval saturates to that for the same primary state.
Contrary to the compact system, the chiral and anti-chiral energy densities decay in $t$ as $t^{-2}$ (for details, see (\ref{eq:late-EDS-inf})) because the constant pieces of those energy densities should be zero.
In this sense, the late-time behavior of the entanglement entropy for the compact system is induced by the leading term of the chiral and anti-chiral energy densities in the large $t$ expansion, while that for the non-compact system is induced by the next-to-leading term.

Furthermore, we discuss the mechanism that causes the difference between the late-time $\Delta S_A$ with $\delta =0$ and that with $\delta \neq 0$ in the non-compact system.
As shown in (\ref{eq:late-EDS-inf}), during the unitary time evolution, i.e., $\delta=0$, the late-time behavior of the chiral and anti-chiral energy densities behaves as $t^{-4}$, while during the unitary time evolution, i.e., $\delta\neq 0$, those do as $t^{-2}$.
This difference between the late-time chiral and anti-chiral energy densities with $\delta=0$ and those with $\delta \neq 0$ may induce that of the late time $\Delta S_A$.


\section{Gravity Duals \label{Sec:Gravity-Duals}}

In this section, we propose gravity duals for the systems considered in the previous sections.
Our proposal is based on the study on the time dependence of the gravity dual in terms of the Euclidean path-integral: we begin with Euclidean gravity corresponding to the Euclidean counterparts of the systems under consideration,  perform the analytic continuation from Euclidean time to Lorentzian, and then, we investigate the time dependence of the gravity dual that is analytically continued.
In this section, we generalize this analytic continuation to that from Euclidean time to complex time, and discuss the gravity dual analytically continued in this manner.

To clearly investigate the properties of the gravity dual of the system under consideration,  we focus on a spinless heavy primary operator $\mathcal O$ with equal chiral and anti-chiral conformal weights $(h_{\mathcal O},\overline{h}_{\mathcal O})=(h_{\mathcal O},h_{\mathcal O})$ and satisfying
\begin{equation}
	\frac{24 h_{\mathcal O}}{c}>1.
\end{equation}
The gravity dual of the system with the insertion of the heavy operator can be approximated by a black hole with the horizon determined by the conformal dimensions of the inserted operator.
Therefore, the intriguing properties of the system during the complex time evolution may be encoded in the behavior of the black hole horizon.

\subsection{Effective Horizon Profile and Relaxation}
We begin with the gravity dual of the Euclidean counterpart of the system with the insertion of a heavy local operator with weights \((h_{\mathcal O},h_{\mathcal O})\).
This Euclidean gravity is approximated by a BTZ black hole \cite{Banados:1992gq}.
For later convenience, we work in Euclidean BTZ coordinates \((r,\xi,\phi)\),
where \(r\) is the radial coordinate, \(\xi\) is the Euclidean time, and \(\phi\sim\phi+2\pi\) is the angular coordinate.
For the static, non-rotating case, the Euclidean BTZ metric reads
\begin{equation}
	\label{eq:BTZ-metric}
	ds_E^2
	=\frac{r^2-r_h^2}{L_{\mathrm{AdS}}^{\,2}}\,d\xi^2
	+\frac{L_{\mathrm{AdS}}^{\,2}}{r^2-r_h^2}\,dr^2
	+r^2\,d\phi^2,
\end{equation}
where \(L_{\mathrm{AdS}}\) is the AdS radius and \(r_h\) is the BTZ horizon radius.
The dimensionless ratio of $r_h $ to $L_{\mathrm{AdS}}$ is related to the conformal weight of the heavy operator as
\begin{equation}
	\label{eq:rh-hc-relation}
	\frac{r_h}{L_{\mathrm{AdS}}}
	=\sqrt{\frac{24 h_{\mathcal{O}}}{c}-1}\,.
\end{equation}
Note that near the boundary \(r\to\infty\), \eqref{eq:BTZ-metric} asymptotes to Euclidean \(\mathrm{AdS}_3\) with radius \(L_{\mathrm{AdS}}\).

Then, to clarify the relation between the black hole and the geometry determined by the chiral and anti-chiral energy densities on the boundary, we map from the BTZ geometry to the Poincaré patch by using the coordinate transformation
\begin{equation}
	\label{eq:map-btz-ads}
	\begin{aligned}
		\tilde{w} &= \sqrt{1-\Bigl(\frac{r_h}{r}\Bigr)^{\!2}}\;
		\exp\!\left[\frac{r_h}{L_{\mathrm{AdS}}}\Bigl(\phi+\frac{i \xi}{L_{\mathrm{AdS}}}\Bigr)\right],\\[2pt]
		\overline{\tilde{w}} &= \sqrt{1-\Bigl(\frac{r_h}{r}\Bigr)^{\!2}}\;
		\exp\!\left[\frac{r_h}{L_{\mathrm{AdS}}}\Bigl(\phi-\frac{i \xi}{L_{\mathrm{AdS}}}\Bigr)\right],\\[2pt]
		\tilde{\nu} &= \frac{r_h}{r}\,\exp\!\left[\frac{r_h}{L_{\mathrm{AdS}}}\,\phi\right].
	\end{aligned}
\end{equation}
Consequently,  the BTZ black hole in \eqref{eq:BTZ-metric} is locally mapped to the pure \(\mathrm{AdS}_3\) in \eqref{eq:pure-ads-3}.

By combining \eqref{TS:BNA-ADS} with \eqref{eq:map-btz-ads}, we obtain a direct map from the Bañados coordinates \((w,\overline{w},\nu)\) to the BTZ coordinates \((r,\xi,\phi)\).
For our purposes, it suffices to consider the asymptotic map in the near-boundary region, \(r\to\infty\).
In this region, the direct map is simplified as 
\begin{equation}
	\label{eq:asymptotic-form-coordinate-transformation}
	e^{2 i\frac{r_h}{L_{\mathrm{AdS}}}\frac{\xi}{L_{\mathrm{AdS}}}}
	\approx \frac{f(w)}{\overline{f}(\overline{w})},\qquad
	e^{2\frac{r_h}{L_{\mathrm{AdS}}}\phi}
	\approx f(w)\,\overline{f}(\overline{w}),\qquad
	r^2 \approx \frac{r_h^2\,f(w)\,\overline{f}(\overline{w})}{\nu^2 f'(w)\overline{f}'(\overline{w})}\,,
\end{equation}
which determines how hypersurfaces of constant BTZ radius \(r\) are embedded in the Bañados coordinates, and in particular allows us to track the position of the horizon \(r=r_h\) as a function of the boundary coordinates.
It should be pointed out that this function is determined by the chiral and anti-chiral energy densities in the holographic CFT system.

To clearly see the profile of the horizon that reflects the properties of the CFT system, we start from the BTZ metric in \((r,\xi,\phi)\), and rewrite the geometry, near the boundary, i.e., $r\rightarrow \infty$, 
in terms of the Bañados boundary coordinates \((w,\overline{w})\).
By exploiting \eqref{eq:asymptotic-form-coordinate-transformation}, we obtain the asymptotic geometry near the boundary as
\begin{equation}
	\left.ds^2\right|_{r \rightarrow \infty}
	=\left(\frac{L_{\mathrm{AdS}}}{r}\right)^2 dr^2
	+r^2\,\frac{f'(w)\overline{f}'(\overline{w})}{f(w)\overline{f}(\overline{w})}\,dw\,d\overline{w}\, .
\end{equation}
It is convenient to introduce the radial variable as
\begin{equation}
	\label{eq:rho-def}
	\rho(r;w,\overline{w})
	:= \left(\frac{L_{\mathrm{AdS}}^2}{r^2}\,
	\frac{f(w)\overline{f}(\overline{w})}{f'(w)\overline{f}'(\overline{w})}\right)^{\frac{1}{2}}.
\end{equation}
After rewriting the geometry with $\rho$, near the boundary, $\rho \rightarrow 0$, the geometry takes the standard Poincar\'e \(\mathrm{AdS}_3\),
\begin{equation}
	ds^2 \left|_{\rho\to 0} \right. = L_{\mathrm{AdS}}^{\,2}\,\frac{d\rho^2+dw\,d\overline{w}}{\rho^2}.
\end{equation}
With this convention, the conformal boundary is located at \(\rho\to 0\) (equivalently \(r\to\infty\)),
and \(\rho\) increases toward the interior.
It should be pointed out that after performing the analytic continuation as $\tau_1=t$ and $\tau_2=it$, we obtain  (\ref{eq:com-geo-und-con}) as the boundary geometry.

Now, we report on the spacetime dependence, reflecting that of the CFT dual, of the black hole horizon.
In the coordinates $(\rho, w ,\overline{w})$, the location of the black hole horizon is
\begin{equation}
	\label{eq:rho-horizon-latecompact}
	\rho(r_{h};w,\overline{w})
	= \left(\frac{L_{\mathrm{AdS}}^2}{r_{h}^2}\,
	\frac{f(w)\overline{f}(\overline{w})}{f'(w)\overline{f}'(\overline{w})}\right)^{\frac{1}{2}}.
\end{equation}
After performing the analytic continuation as $\tau_1=t$ and $\tau_2=it$, we can obtain the spacetime-dependence of the black hole horizon in the gravity dual of the compact system under consideration.

 We begin by looking at the spacetime dependence of the black hole horizon in the gravity dual of the compact system investigated in Section \ref{sec:2d-com-hol}.
First, we describe the late-time behavior of the black hole horizon.
If we expand the black hole horizon with respect to $\delta t \gg 1$, then at the leading order in this expansion, we obtain the late-time behavior of the black hole horizon as 
\begin{equation} \label{eq:late-time-com-hol}
	\rho (r_{h};w,\overline{w})= \left(\frac{L_{\mathrm{AdS}}}{r_{h}}\right)^2 \frac{L}{2 \pi} \left[1+O\!\left((\delta t)^{-1}\right)\right].
\end{equation}
Thus, during the complex time evolution, the black hole horizon saturates to the static and homogeneous one.
By exploiting the relation in (\ref{eq:rh-hc-relation}), we can see that the late time behavior of the black hole is approximated as the static one of the black hole which is dual of the primary state with large conformal dimensions, $(h_{\mathcal{O}}, h_{\mathcal{O}})$.

\begin{figure}[htbp]
	\centering
	\subfloat[$\delta = 0.$]{\includegraphics[width=.3\columnwidth]{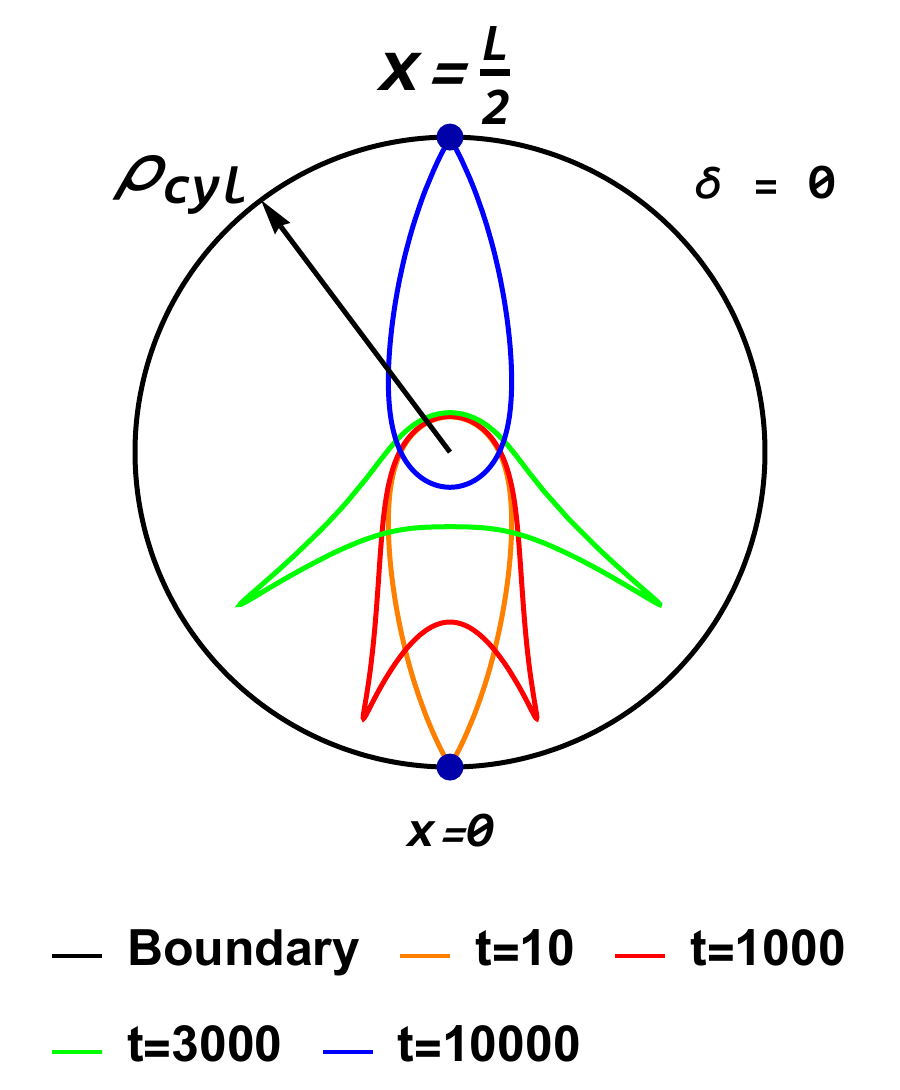}}\hspace{5pt}
	\subfloat[$\delta = 0.1.$]{\includegraphics[width=.3\columnwidth]{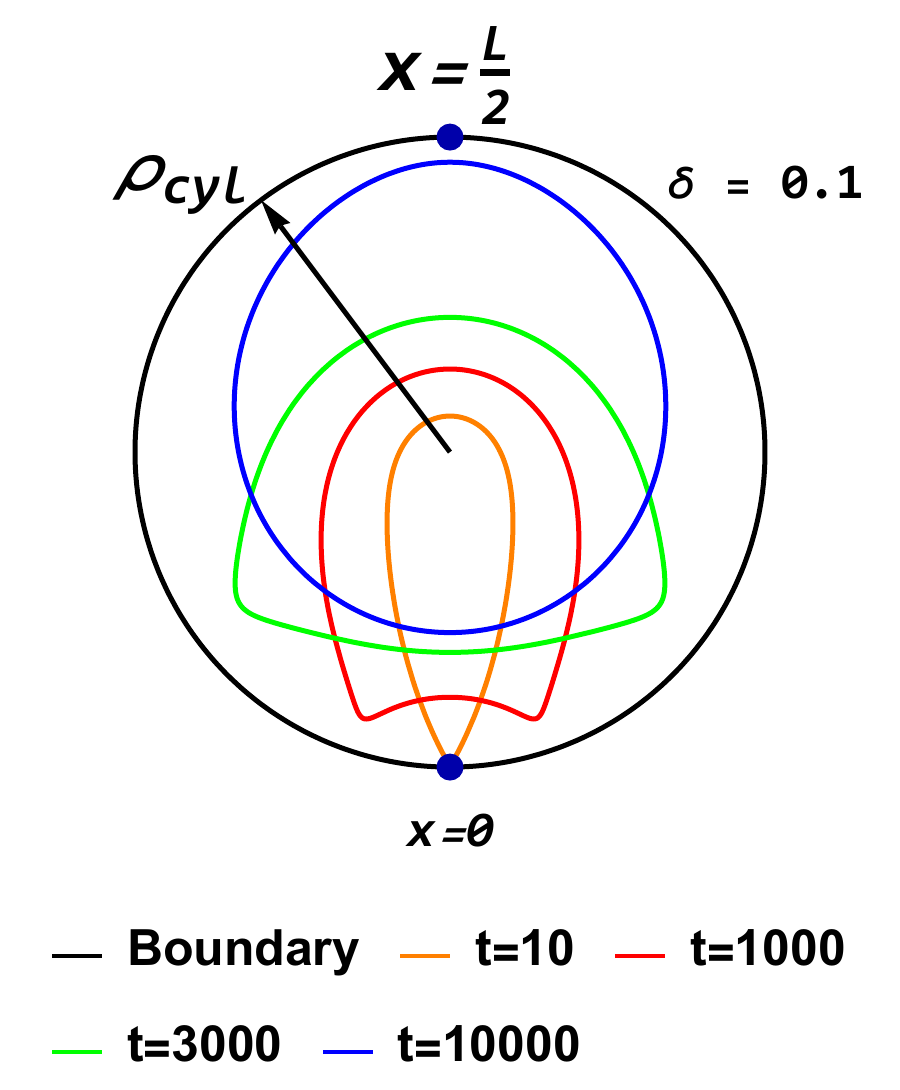}}\hspace{5pt}
	\subfloat[$\delta =  1.$]{\includegraphics[width=.3\columnwidth]{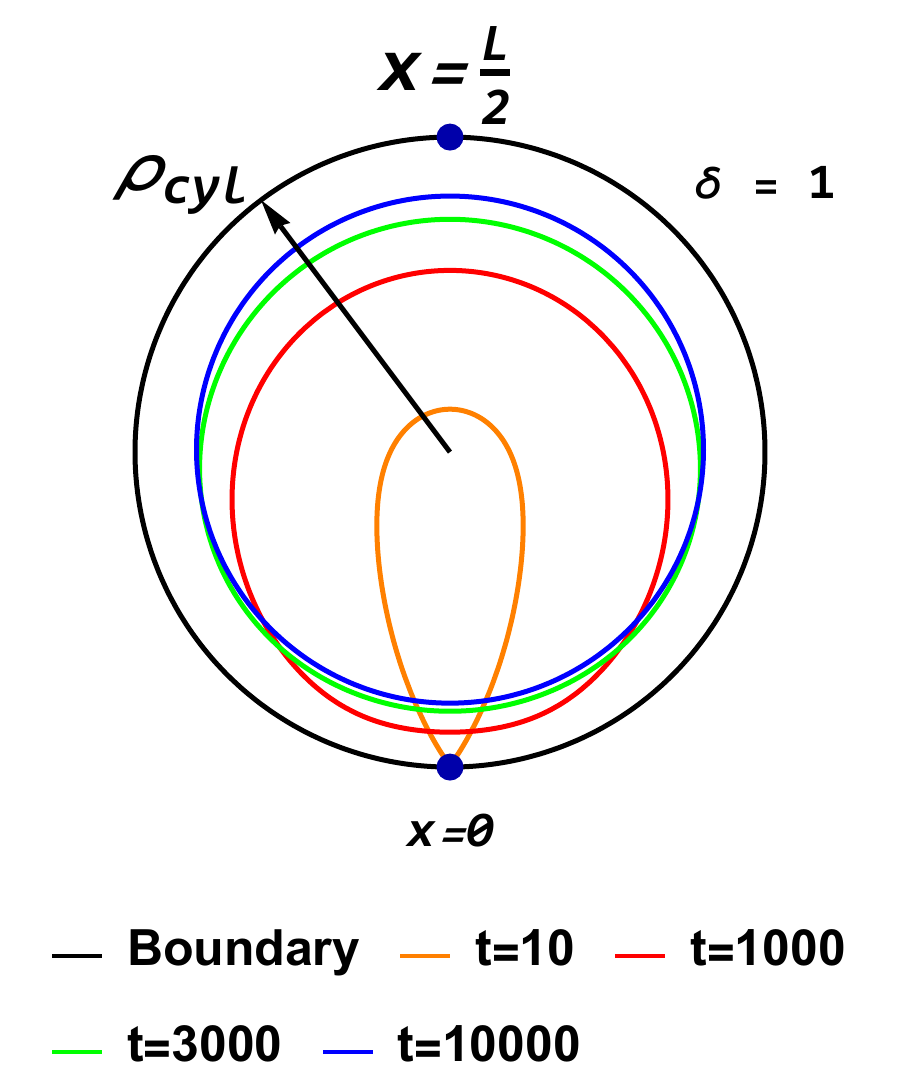}}\hspace{5pt}
	\caption{
		Profiles of the horizon on a compact circle.
		In panels~(a)–(c), we show the horizon position \(\rho^{cyl}_{h_\mathcal{O}}(X)\) as a function of the angular coordinate \(X\) at several fixed times \(t\),
		for \(\beta = 20\),  \(\frac{r_{h}}{L_{\mathrm{AdS}}}=10\), and \(L=20000\),
		with \(\delta = 0\), \(0.1\), and \(1\), respectively.
		In this figure, we assume that the local operator is inserted at \(x=0\). 
	}
	\label{Fig:Horizon-1}
\end{figure}

Then, let us report the spacetime dependence of the black hole horizon.
To easily see the spacetime dependence of the black hole horizon in the compact system, we introduce a new radial variable as 
\begin{equation}
	\label{eq:rho-def-cyl}
	\rho^{cyl}(r;w,\overline{w})
	:=\frac{L}{2\pi}\,
	\left(\frac{\displaystyle \frac{\pi}{2}-\tan^{-1}\!\left[
		\left(\frac{L_{\mathrm{AdS}}^2}{r^2}\,
		\frac{f(w)\overline{f}(\overline{w})}{f'(w)\overline{f}'(\overline{w})}\right)^{1/2}
		\right]}
	{\displaystyle \frac{\pi}{2}}\right),
\end{equation}
By using this variable, the spacetime dependence of the black hole horizon is expressed as 
\begin{equation}
	\label{eq:rho-horizon-cyl}
	\rho^{cyl}(r_{h};w,\overline{w})
	:=\frac{L}{2\pi}\,
	\left(\frac{\displaystyle \frac{\pi}{2}-\tan^{-1}\!\left[
		\left(\frac{L_{\mathrm{AdS}}^2}{r_{h}^2}\,
		\frac{f(w)\overline{f}(\overline{w})}{f'(w)\overline{f}'(\overline{w})}\right)^{1/2}
		\right]}
	{\displaystyle \frac{\pi}{2}}\right).
\end{equation}
In Fig. \ref{Fig:Horizon-1}, we show the position dependence of the horizon defined as \eqref{eq:rho-horizon-cyl} for various times with $\delta=0, 0.1$ and $1$. In this figure, we insert the local operator at $x=0$.
From this figure, we can observe the following.
With $\delta=0$, a horizon peak emerges at the insertion point of the local operator, and then splits into two peaks during the unitary time evolution. 
Subsequently, these two peaks form a single peak around $x=\f{L}{2}$. 
These two peaks survive during the time evolution and periodically move in time with the period of $L$.
This is because the chiral and anti-chiral energy densities are the periodic functions in time with the period of $L$.
For $\delta \neq 0$, a single peak initially emerges at the insertion point of the local operator, and then splits into two during the non-unitary time evolution.
In contrast to the unitary time evolution, these peaks become rounder and rounder during the non-unitary time evolution.
Eventually, the shape of the horizon approximately becomes round.

From the late-time behavior in (\ref{eq:late-time-com-hol}) and observations from Fig. \ref{Fig:Horizon-1}, we can see that during the complex time, the time dependence of the black horizon exhibits the relaxation from the black hole with the inhomogeneous horizon to that with the homogeneous one determined by the conformal dimensions of the inserted operator.
 On the CFT side, the time evolution of the black hole horizon, during the complex time evolution, can describe the relaxation from the state with a local excitation induced by the inserted heavy operator to the thermal state. 

Then, let us present the spacetime dependence of the black hole horizon in the non-compact system, i.e., $L\rightarrow \infty$.
By using the relation between the chiral and anti-chiral energy densities and the holomorphic and anti-holomorphic functions in (\ref{eq:schwarzian}), for non-compact case, we can simplify the location of the black hole horizon as  
\begin{equation}
	\label{eq:rho-horizon-latenoncompact}
\rho^{L \rightarrow \infty}
 (r_{h};w,\overline{w})
	= \left(\frac{L_{\mathrm{AdS}}^2}{r_{h}^2}\,
	\frac{f(w)\overline{f}(\overline{w})}{f'(w)\overline{f}'(\overline{w})}\right)^{\frac{1}{2}}
	=\left(\frac{L_{\mathrm{AdS}}}{r_{h}}\right)\left(\frac{h_{\mathcal{O}}^2}{\left\langle T\left(w\right)\right\rangle \left\langle\overline{T}\left(\overline{w}\right)\right\rangle}\right)^{\frac{1}{4}}.
\end{equation}
After performing the analytic continuation as $\tau_1=t$ and  $\tau_2=it$, we can obtain the spacetime dependence of the black hole horizon as the analytic continued counterpart of (\ref{eq:rho-horizon-latenoncompact}).
From (\ref{eq:rho-horizon-latenoncompact}), we can see that the spacetime dependence of the black hole horizon is determined by that of the chiral and anti-chiral energy densities.
First, we present the late-time behavior of the horizon during the complex time evolution, i.e., \(\delta \neq 0\).
This may reflect the properties of the non-compact system investigated in Section \ref{sec:2dhCFT-inf}.
We substitute \eqref{eq:EDS-inf} into \eqref{eq:rho-horizon-latenoncompact}, expand it with respect to $L \gg 1$ , and then we define the black hole horizon in the non-compact system as the leading term in this expansion.
Furthermore, we expand this black hole horizon with respect to $t\gg 1$, and obtain the late-time behavior of the black hole horizon in the non-compact system as the leading term in this expansion,
\begin{equation}
	\rho^{L \rightarrow \infty} (r_{h};w,\overline{w})\big|_{\delta>0}
	= \frac{L_{\mathrm{AdS}}}{r_{h}}\left[\frac{1+\delta^2}{2 \delta}\, t+O\!\left(t^{0}\right)\right].
\end{equation}
Thus, in the non-compact system, the late-time behavior of the black hole horizon spatially homogeneous, and grows linearly with \(t\).
Contrary to the compact system, the late-time behavior of the horizon depends on $\delta$.
In other words, the late-time behavior of the black hole horizon is controlled by the non-unitary process, $e^{-H\delta t }$.
This is minimized at $\delta=1$ as 
\be
\rho^{L \rightarrow \infty} (r_{h};w,\overline{w})|_{\delta=1}=\frac{L_{\mathrm{AdS}}}{r_{h}} t.
\ee
At the leading order in the large $\delta$ expansion, i.e., $\delta \gg 1$, the late-time behavior of the black hole horizon is proportional to $\delta$.

\begin{figure}[htbp]
	\centering
	\subfloat[$\delta = 0.$]{\includegraphics[width=.3\columnwidth]{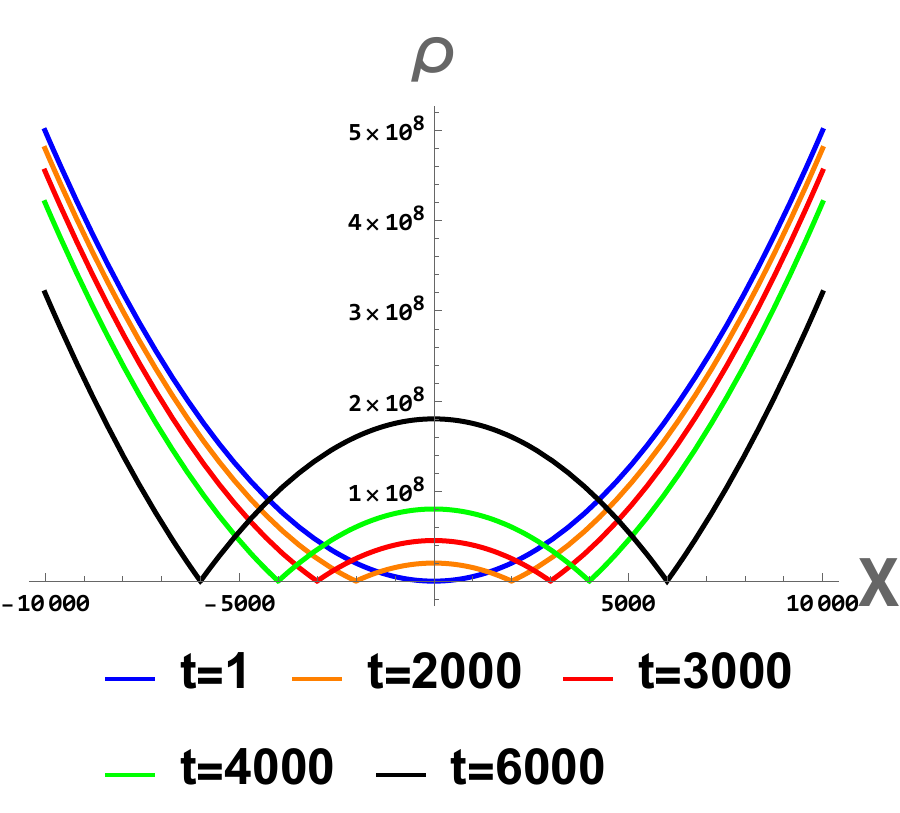}}\hspace{5pt}
	\subfloat[$\delta = 0.1.$]{\includegraphics[width=.3\columnwidth]{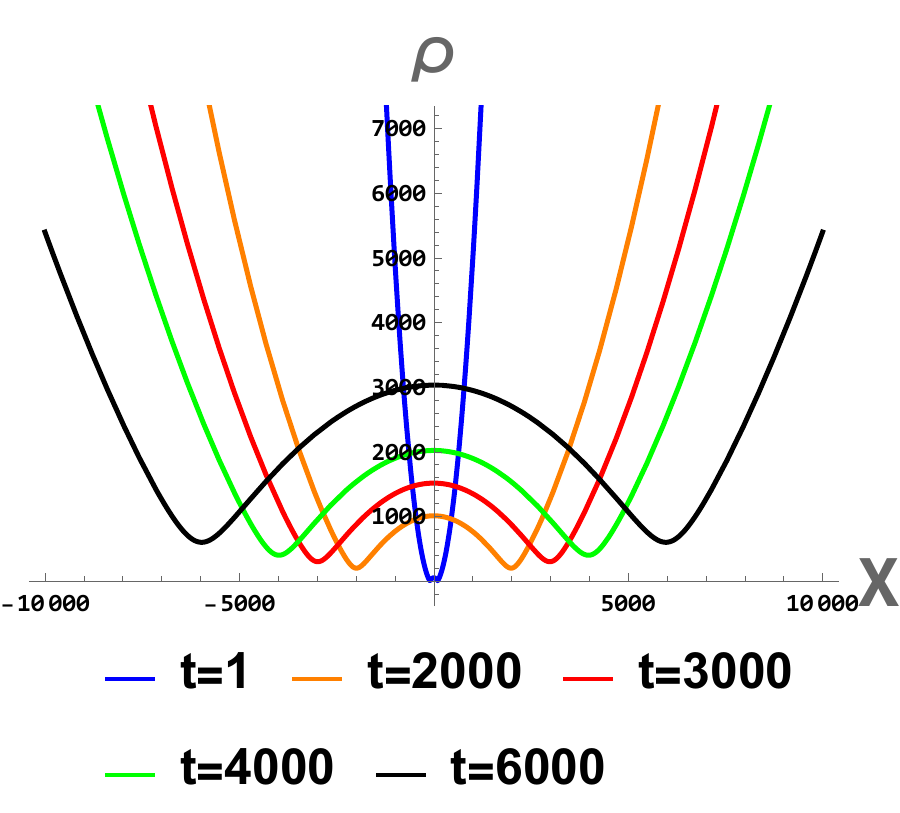}}\hspace{5pt}
	\subfloat[$\delta = 1.$]{\includegraphics[width=.3\columnwidth]{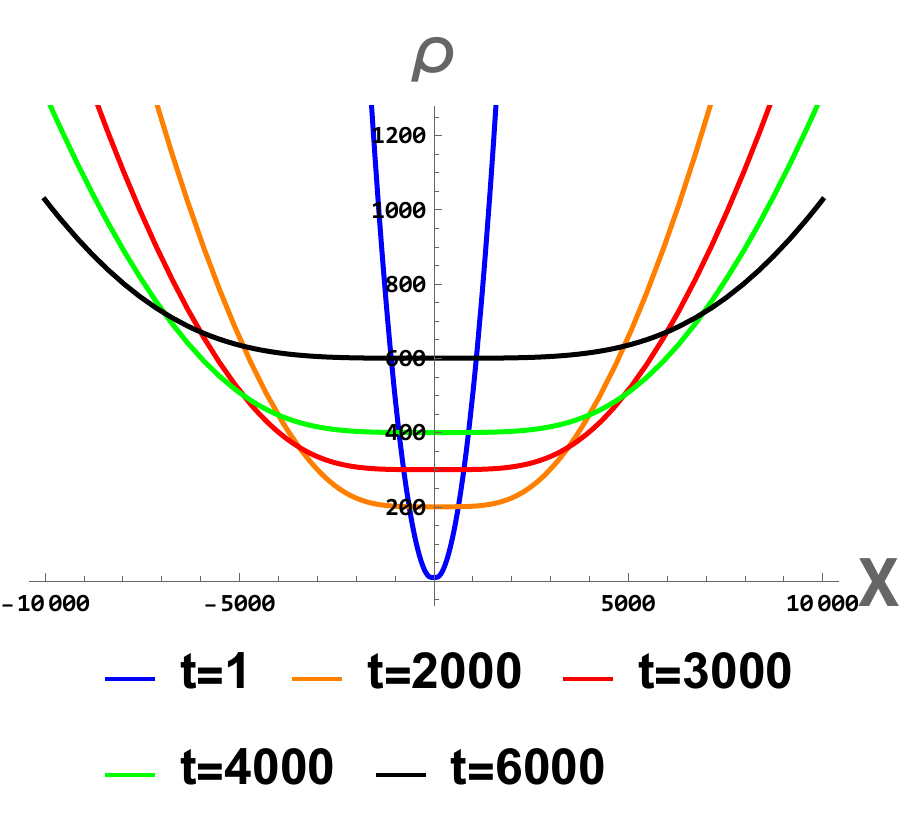}}\hspace{5pt}
	\caption{
		Horizon profiles in the planar geometry.
		Panels~(a)–(c) plot the horizon position $\rho_h(X)$ at several fixed times $t$,
		for holographic CFT data \(\beta = 20\), \(\frac{r_{h}}{L_{\mathrm{AdS}}}=10\), and \(x=0\).
	}
	\label{Fig:Horizon-2}
\end{figure}

In Fig.~\ref{Fig:Horizon-2}, we show the position dependence of (\ref{eq:rho-horizon-latenoncompact}) for the non-compact system with various $t$ and $\delta$.
For all $\delta$, we can observe that the black hole with $t=1$ has a single peak at $X=0$.
For $\delta =0$ and $0.1$, during the time evolution, this single peak splits into two peaks, and then they propagate far away from the insertion point of the local operator, i.e., $x=0$.
Furthermore, we can observe that during the time evolution with $\delta=0$, the location of peaks along the $\rho$ direction does not change, while it moves far away from the boundary during the time evolution with $\delta=0.1$.
From Panel (c), we can observe that during the time evolution with $\delta =1$, the location, along $\rho$, of the black hole horizon moves away from the boundary and its shape flattens according to the increase of the complex time. 

\section{Discussions and Future Directions \label{Sec:Discussions-and-Future-Directions}}
We will close this paper with the discussions on our findings and future directions.
We found that the subsystems of the compact system with the insertion of the primary operator evolve in the complex time to those of the primary state corresponding to the inserted operator.
This suggests that the complex time evolution relaxes the compact system with the insertion of the primary operator to the primary state with the same conformal dimensions as those of the inserted operator.
We will consider the intuitive mechanism that describes this relaxation process.
Rewrite the system in (\ref{eq:system-with-LO}) as
\be
\ket{\Psi_{\mathcal{O}}(t)}=\mathcal{N}_{\mathcal{O}}(t)e^{-itH} e^{-(\delta t + \beta) H} \mathcal{O} (x)e^{(\delta t + \beta) H} \ket{0}
=\mathcal{N}_{\mathcal{O}}(t)e^{-itH}  \mathcal{O}_{\text{H}} (x,-\delta t -\beta) \ket{0},
\ee
where we assume that $H\ket{0}=0$, and $\mathcal{O}_{\text{H}} (x,-\delta t -\beta)$ is defined as $\mathcal{O}_{\text{H}} (x,-\delta t -\beta)=e^{-(\delta t + \beta) H} \mathcal{O} (x)e^{(\delta t + \beta) H} $.

Thus, the Euclidean time evolution operator, induced by the imaginary part of the complex time, acts on the local operator as the translation along the Euclidean time direction.
Taking $t$ to be infinite can be considered as pushing the local operator to minus infinity along the Euclidean time direction.
This may correspond to the replacement of the vacuum state with the corresponding primary state. 
In this manner, the non-unitary process may evolve the vacuum state to the primary state corresponding to the inserted operator.
Based on this intuitive mechanism, we will discuss a few of future direction.

{\it  CFT Hamiltonian on the curved background:} 
One simple generalization of this paper could be replacing the uniform Hamiltonian with a curved background Hamiltonian such as the SSD Hamiltonian \cite{Tada_2015,Ishibashi_2015,Ishibashi_2016,wen2018floquet,Fan_2020,Fan_2021,PhysRevResearch.3.023044,PhysRevB.102.205125,2022Wen,Fang:2025rie,2021Nozaki,MacCormack_2019,2023arXiv230208009G,2023arXiv231019376N,2023JHEP...03..101G,Miyata:2024gvr,Bai:2024azk,Li:2025rzl,Erdmenger:2025chu,Das:2023xaw,Das:2024lra,Das:2025wjo,2024Mezei,deBoer:2023lrd,Erdmenger:2024xmj,Das:2024vqe,Lapierre:2024lga,Malvimat:2024vhr,Lapierre:2019rwj,Lapierre:2020ftq,Lapierre:2020roc,Mo:2025imx,Mao2024,Bernamonti:2024fgx,2023_OBC,Das:2022jrr,bai2025spatially}.
Changing the Hamiltonian corresponds to changing the direction where the Euclidean time evolution operator, induced by the imaginary part of the complex time, pushes the inserted operator. 
This may yield to the relaxation to the states other than primary states, such as the descendant states.\\

{\it Other measurements:}
As explained in Section \ref{eq:Interpretation-as-measurement}, the Euclidean time evolution can be thought of as the non-unitary process induced by the post-selected projective measurement.
If we replace the Hamiltonian for the Euclidean time evolution of $M$ defined in (\ref{eq:Sigle-step-process}) with other Hamiltonians, we can generalize the system under consideration to Floquet systems with other measurements than the one considered in this paper.
This replacement of the Euclidean time evolution may provide us with the late-time state with a richer entanglement structure. 
Furthermore, such Floquet systems may lead to the measurement-induced phase transition \cite{Lapierre_2025_cft,Lapierre:2025wya,Mo:2025imx}.\\

{\bf Gravity dual of other complex time evolutions:} 
In this paper, we proposed and discussed the gravity dual of the relaxation from the vacuum state to the primary one.
In the case of the heavy local operator, i.e., $\f{24 h_{\mathcal{O}}}{c}>1$, this gravity dual is the time evolution from the inhomogeneous black hole to the homogeneous one.
For the large conformal dimensions, the non-unitary process, under consideration, is reflected by the dynamical behavior of the horizon, i.e, the relaxation to the homogeneous one.
In the case of the light local operator, i.e., $\f{24 h_{\mathcal{O}}}{c}<1$, the dual geometry is no longer the black hole.
Therefore, another object in gravity dual may reflect the time evolution to the steady state, i.e., the primary state.
Probably, as in \cite{Nozaki_2013}, the evolution to the primary state may be reflected by the motion of the local object.
It would be interesting to investigate the gravity dual of the evolution to the primary state.\\

{\it Gravity Dual:} 
It would be interesting to propose the gravity dual of other complex time evolutions, such as global quenches.
The non-unitary processes induced by the complex time evolution may be reflected by some relaxation processes of the gravity dual.
Therefore, such gravity duals may be explored as the non-unitary realization of black hole formation and evaporation.

\section*{Acknowledgment}
We would like to thank Hyunsoo Ha, Xinyu Sun, Mingrui Li, Farzad Omidi and Yu-Xuan Zhang for very helpful discussions.
M.N.~is supported by funds from the University of Chinese Academy of Sciences (UCAS), funds from the Kavli
Institute for Theoretical Sciences (KITS). M.T.T. was supported by an 
appointment to the YST Program at the APCTP through the Science and 
Technology Promotion Fund and Lottery Fund of the Korean Government, and also by the Korean Local Governments -
Gyeongsangbuk-do Province and Pohang City. M.T.T. also acknowledges funding from the European Research
Council (ERC) through Advanced grant QUEST (Grant Agreement No. 101096208).
X.W. is supported by a startup at Georgia Institute of Technology.

\bibliographystyle{ieeetr}
\bibliography{reference.bib}

\appendix

\section{Uniformization and cross ratios for $n=2$}

\subsection{Uniformization map and definition of $(\eta_2,\overline{\eta}_2)$}
\label{app:uniformization}

Here we summarize the geometric ingredients needed to evaluate the second R\'enyi
entropy for a single interval $A=[X_2,X_1]$ on a circle of circumference $L$.
For the standard background on the replica construction and twist operators, see
Refs.~\cite{Calabrese_2004,2009JPhA...42X4005C}; for analytic continuation in
local-operator setups, see e.g. Ref.~\cite{Caputa:2014vaa}.

We map the Euclidean cylinder to the complex plane by
\begin{equation}
	z=e^{\frac{2\pi}{L}w},\qquad \overline{z}=e^{\frac{2\pi}{L}\overline{w}}.
	\label{eq:app-zmap}
\end{equation}
The endpoints of the interval (located at $\tau=0$) are
\begin{equation}
	z_{X_i}=e^{\frac{2\pi i}{L}X_i},\qquad \overline{z}_{X_i}=e^{-\frac{2\pi i}{L}X_i},
	\qquad (i=1,2).
\end{equation}
Let $z_{1,2}$ and $\overline{z}_{1,2}$ denote the images of the operator insertions
defined in Eq.~\eqref{eq:insert-positions}.

For $n=2$, the replica surface $\Sigma_2$ is a two-sheeted cover of the $z$-plane
branched at $z_{X_1}$ and $z_{X_2}$. It is uniformized by the square-root map
\begin{equation}
	\zeta(z)=\sqrt{\frac{z-z_{X_2}}{z_{X_1}-z}},
	\qquad
	\overline{\zeta}(\overline{z})=\sqrt{\frac{\overline{z}-\overline{z}_{X_2}}{\overline{z}_{X_1}-\overline{z}}}.
	\label{eq:app-zeta}
\end{equation}
The two sheets correspond to the two signs of the square root, $\zeta\leftrightarrow -\zeta$
(and similarly in the barred sector). Define
\begin{equation}
	\zeta_1\equiv \zeta(z_1),\quad \zeta_2\equiv \zeta(z_2),
	\qquad
	\overline{\zeta}_1\equiv \overline{\zeta}(\overline{z}_1),\quad \overline{\zeta}_2\equiv \overline{\zeta}(\overline{z}_2).
\end{equation}
A convenient choice of the four points on the uniformized plane is
\begin{equation}
	(\zeta_1,\zeta_2,\zeta_3,\zeta_4)=(\zeta_1,\zeta_2,-\zeta_1,-\zeta_2),
	\qquad
	(\overline{\zeta}_1,\overline{\zeta}_2,\overline{\zeta}_3,\overline{\zeta}_4)=(\overline{\zeta}_1,\overline{\zeta}_2,-\overline{\zeta}_1,-\overline{\zeta}_2).
	\label{eq:app-4pts}
\end{equation}
The corresponding cross ratios are
\begin{equation}
	\eta_2
	=\frac{(\zeta_1-\zeta_2)(\zeta_3-\zeta_4)}{(\zeta_1-\zeta_3)(\zeta_2-\zeta_4)}
	=-\frac{(\zeta_1-\zeta_2)^2}{4\,\zeta_1\,\zeta_2},
	\label{eq:app-eta2}
\end{equation}
\begin{equation}
	\overline{\eta}_2
	=\frac{(\overline{\zeta}_1-\overline{\zeta}_2)(\overline{\zeta}_3-\overline{\zeta}_4)}{(\overline{\zeta}_1-\overline{\zeta}_3)(\overline{\zeta}_2-\overline{\zeta}_4)}
	=-\frac{(\overline{\zeta}_1-\overline{\zeta}_2)^2}{4\,\overline{\zeta}_1\,\overline{\zeta}_2}.
	\label{eq:app-etabar2}
\end{equation}
We will also use the identity
\begin{equation}
	1-\eta_2=\frac{(\zeta_1+\zeta_2)^2}{4\,\zeta_1\,\zeta_2},
	\qquad
	1-\overline{\eta}_2=\frac{(\overline{\zeta}_1+\overline{\zeta}_2)^2}{4\,\overline{\zeta}_1\,\overline{\zeta}_2}.
	\label{eq:app-1minus-eta}
\end{equation}
Because $\zeta$ is multi-valued, analytic continuation may move the correlator
between different analytic branches (equivalently, different OPE channels). In the
main text we always choose the branch continuously along the prescribed contour,
as standard in local-operator analyses \cite{Caputa:2014vaa}.

\subsection{Analytic continuation and late-time limit}
\label{app:eta}

We implement the analytic continuation used in Sec.~\ref{sec:SREE},
\begin{equation}
	\tau_1=t,\qquad \tau_2=i t.
	\label{eq:app-analytic}
\end{equation}
Substituting Eq.~\eqref{eq:app-analytic} into Eq.~\eqref{eq:insert-positions}, one finds
\begin{equation}
	w_1=-(\beta+\delta t+i t)+ix,\qquad
	w_2=(\beta+\delta t-i t)+ix,
\end{equation}
and therefore, using Eq.~\eqref{eq:app-zmap},
\begin{equation}
	z_1=\exp\!\left[-\frac{2\pi}{L}(\beta+\delta t)\right]
	\exp\!\left[-\frac{2\pi i}{L}(t-x)\right],
	\qquad
	z_2=\exp\!\left[+\frac{2\pi}{L}(\beta+\delta t)\right]
	\exp\!\left[-\frac{2\pi i}{L}(t-x)\right],
	\label{eq:app-z1z2}
\end{equation}
with $\overline{z}_{1,2}$ obtained by $x\to -x$ in the phase.

We now derive the late-time plateau used in Eq.~\eqref{eq:eta-late}.
Consider the regime
\begin{equation}
	\frac{2\pi}{L}(\beta+\delta t)\gg 1,
	\label{eq:app-late}
\end{equation}
so that $|z_1|\to 0$ and $|z_2|\to\infty$ (up to an overall phase) from
Eq.~\eqref{eq:app-z1z2}. In the uniformization map \eqref{eq:app-zeta}, define
\begin{equation}
	u(z)\equiv \frac{z-z_{X_2}}{z_{X_1}-z},\qquad \zeta(z)=\sqrt{u(z)}.
\end{equation}
Then, in the limit \eqref{eq:app-late},
\begin{equation}
	u(z_1)\longrightarrow -\frac{z_{X_2}}{z_{X_1}}
	=-e^{-\frac{2\pi i}{L}(X_1-X_2)},
	\qquad
	u(z_2)\longrightarrow -1.
\end{equation}
Taking the principal square roots gives
\begin{equation}
	\zeta_1=\zeta(z_1)\longrightarrow i\,e^{-\frac{i\pi}{L}(X_1-X_2)},
	\qquad
	\zeta_2=\zeta(z_2)\longrightarrow i.
\end{equation}
Substituting these into Eq.~\eqref{eq:app-eta2}, we obtain
\begin{equation}
	\eta_2 \longrightarrow
	-\frac{\big(i e^{-\frac{i\pi}{L}(X_1-X_2)}-i\big)^2}{4\,(i e^{-\frac{i\pi}{L}(X_1-X_2)})\,i}
	=\sin^2\!\left[\frac{\pi(X_1-X_2)}{2L}\right],
	\label{eq:app-eta-late}
\end{equation}
and hence, from Eq.~\eqref{eq:app-1minus-eta},
\begin{equation}
	1-\eta_2 \longrightarrow
	\cos^2\!\left[\frac{\pi(X_1-X_2)}{2L}\right]
	=1-\sin^2\!\left[\frac{\pi(X_1-X_2)}{2L}\right].
	\label{eq:app-1minus-eta-late}
\end{equation}
\subsection{Reduced correlators for $O_1$ and $O_2$ in the $c=1$ free boson}
\label{app:O1O2}

In Sec.~\ref{sec:SREE} we investigated the complex time dependence of $\Delta S^{(2)}_A$ for the vacuum state system with the insertion of the holomorphic current $\mathcal O=i\partial\phi$, and then found that it evolves to the primary state with the same conformal dimension as the inserted operator.
Here, we will investigate the same non-equilibrium phenomena with the inserted operator replaced with others.
We choose the inserted operators to be
\begin{equation}
	O_1=e^{\frac{i}{2}\phi},\qquad
	O_2=\frac{1}{\sqrt{2}}\left(e^{\frac{i}{2}\phi}+e^{-\frac{i}{2}\phi}\right),
\end{equation}
which have conformal weights
\begin{equation}
	(h_{O_1},\overline{h}_{O_1})=(h_{O_2},\overline{h}_{O_2})=\left(\frac18,\frac18\right).
\end{equation}

Using Wick's contraction, we obtain the reduced correlators as 
\begin{align}
	G_{1}(\eta_2,\overline{\eta}_2)
	&=\frac{1}{\sqrt{|\,\eta_2\,|\,|\,1-\eta_2\,|}}
	=\bigl[\eta_2\overline{\eta}_2(1-\eta_2)(1-\overline{\eta}_2)\bigr]^{-1/4},
	\label{eq:app-G1}\\[4pt]
	G_{2}(\eta_2,\overline{\eta}_2)
	&=\frac{|\,\eta_2\,|+1+|\,1-\eta_2\,|}{2}\,G_{1}(\eta_2,\overline{\eta}_2)\\
	&=\frac{1+\sqrt{\eta_2\overline{\eta}_2}+\sqrt{(1-\eta_2)(1-\overline{\eta}_2)}}{2}\,
	\bigl[\eta_2\overline{\eta}_2(1-\eta_2)(1-\overline{\eta}_2)\bigr]^{-1/4}.
	\label{eq:app-G2}
\end{align}
This expression is for the Euclidean path-integral.
After analytic continuation, we treat $\eta_2$ and $\overline{\eta}_2$ as independent and  choose the branch such that $\Delta S^{(2)}_A$ becomes a continuous function of time.

Substituting \eqref{eq:app-G1} and \eqref{eq:app-G2} into \eqref{eq:DeltaS2-master}
with $h_{\mathcal{O}}=\overline{h}_{\mathcal{O}}=\frac18$, we obtain the late time second R\'enyi entanglement entropies.
In the late time region, $\Delta S^{(2)}_{A}$ for $\mathcal{O}_1$ vanishes as
\begin{equation}
	\Delta S^{(2)}_{A}\Big|_{O_1}
	=-\log\!\left\{\big[\eta_2(1-\eta_2)\big]^{1/4}\big[\overline{\eta}_2(1-\overline{\eta}_2)\big]^{1/4}\,G_1(\eta_2,\overline{\eta}_2)\right\}
	=0.
\end{equation}
Furthermore, in the late time region, $\Delta S^{(2)}_{A}$ for $\mathcal{O}_2$ vanishes as
\begin{equation}
	\begin{aligned}
	\Delta S^{(2)}_{A}\Big|_{O_2}
	&=-\log\!\left\{\big[\eta_2(1-\eta_2)\big]^{1/4}\big[\overline{\eta}_2(1-\overline{\eta}_2)\big]^{1/4}\,G_2(\eta_2,\overline{\eta}_2)\right\}\\
	&=-\log\!\left[\frac{1+\sqrt{\eta_2\overline{\eta}_2}+\sqrt{(1-\eta_2)(1-\overline{\eta}_2)}}{2}\right]=0.
	\end{aligned}
\end{equation}  
The late time behavior of $\Delta S^{(2)}_A$ for both $\mathcal{O}_1$ and $\mathcal{O}_2$ matches that for the primary states with the same conformal dimensions of the inserted local operators.

\section{2d Ising CFT on Infinite Space \label{sec:2d-ISing-CFT}}
\begin{figure}[ht]
\centering
        \begin{tikzpicture}
        
        \node[inner sep=0pt] (russell) at (-120pt,-20pt)
        {\includegraphics[width=3.0in]{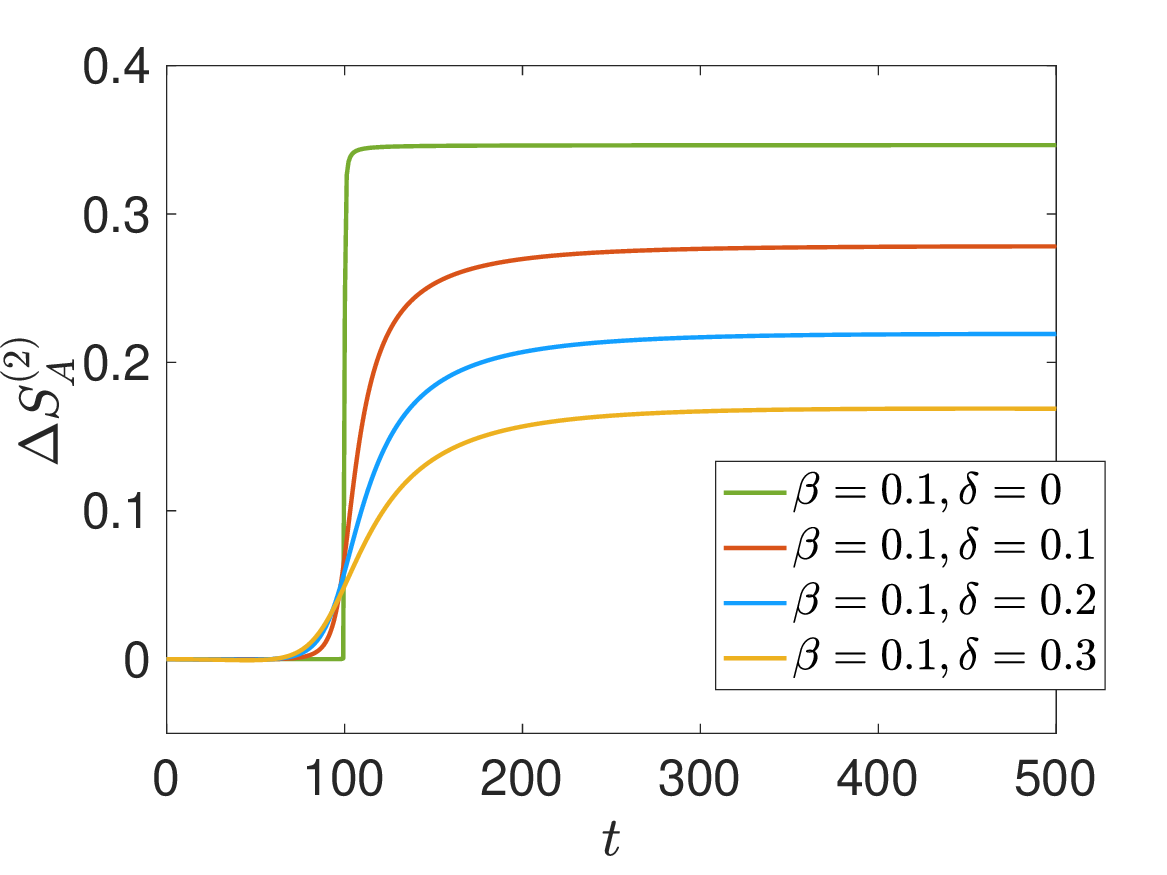}};

         \node[inner sep=0pt] (russell) at (95pt,-20pt)
        {\includegraphics[width=3.0in]{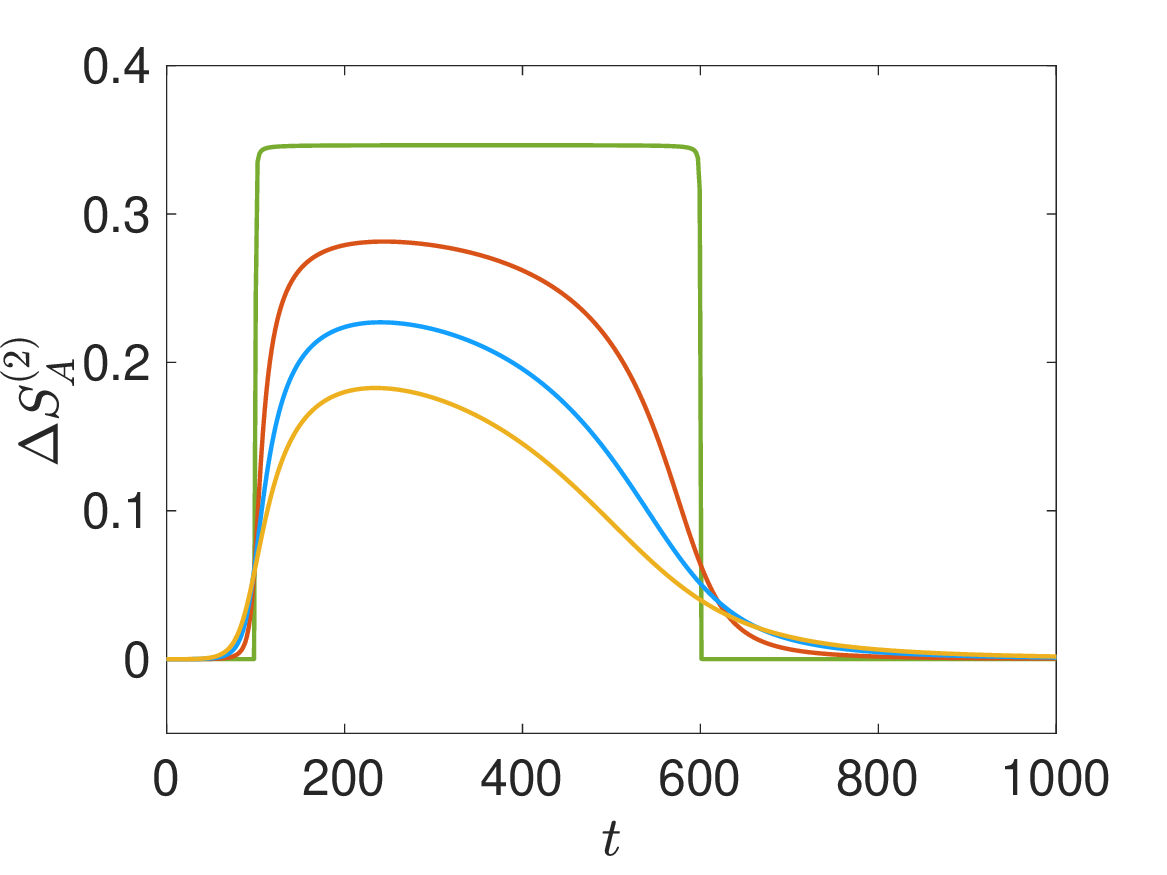}};       
\small
\draw [dashed][red][line width=1.0] (-196pt,33.5pt) -- (-166pt,33.5pt);

\draw (-215pt,33.5pt) node  {$\textcolor{red}
{\frac{1}{2}\log 2}$};

\end{tikzpicture}
	\caption{
    Second Rényi entropy evolution in Ising CFT after a local quantum quench, where the total system is defined on $(-\infty,+\infty)$.
Left: 
The local $\sigma$ operator is inserted at $x=-l=-100$, and $A=[0,+\infty)$.
For $\delta=0$, i.e., there is only a real-time evolution, we have $\Delta S_A^{(2)}=\log d_\sigma=\log \sqrt{2}=\frac{1}{2}\log (2)\simeq 0.3466$ in the late time regime $t\gg l$. In a complex time evolution with $\delta>0$,
the steady value of $\Delta S_A^{(2)}$ depends on $\delta$ according to \eqref{S2} and \eqref{zzbar_longTime}. Right: We take $l=100$, and a finite subsystem $A=[0,l_A]=[0,500]$.
In the late time regime $\Delta S_A^{(2)}$ gradually decays to zero.
    }\label{Fig:Ising}
\end{figure}

In this appendix, we study the time evolution of the Rényi entanglement entropy in another simple example -- the Ising conformal field theory -- on infinite space within the setup of Sec.~\ref{Sec:Local-Operator-Quenches}.
There are three primary fields in Ising CFT, $\mathbb I$, $\epsilon$, and $\sigma$.
Among them, the only nontrivial primary field with quantum dimension greater than one is the 
$\sigma$ field, which has quantum dimension $d_\sigma=\sqrt 2$.
Its conformal dimension is $h_\sigma=\overline{h}_\sigma=1/16$.

We insert the $\sigma$ field at $x=-l$, and study the second Rényi entropy of a subsystem.
Unlike the free-boson case studied in the main text, the primary state associated with the $\sigma$ field exhibits the same Rényi entropy as the ground state \cite{2012JSMTE..01..016I}. As a result, the scaling dimension of the $\sigma$ field does not manifest in the late-time regime of the complex time evolution.

Here we consider two choices of $A$, with $A=[0,+\infty)$ and $A=[0,l_A]$ respectively.
The total system is defined over $(-\infty,+\infty)$.
Following the procedure in Sec.\ref{Sec:Local-Operator-Quenches}, one can obtain the change of second Rényi entropy after a local quench as
\be
\label{S2}
\begin{split}
\Delta S_A^{(2)}
=&
-\log \left(\frac{1}{2}
\sqrt{1+\sqrt{1-\eta_2}} \cdot \sqrt{1+\sqrt{1-\overline{\eta}_2}}+ \frac{1}{2}\sqrt{1-\sqrt{1-\eta_2}} \cdot \sqrt{1-\sqrt{1-\overline{\eta}_2}}
\right)
\end{split}
\ee
where $w_i$ and $\overline{w}_i$ are defined in Eq.~\eqref{eq:insert-positions},
and we map the replica surface $\Sigma_2$ to a complex $z$-plane via the conformal mapping 
\be\label{map:w_z2}
w=z^2.
\ee
The cross ratios $\eta_2$ and $\overline{\eta}_2$ in \eqref{S2} are defined as
$
\eta_2=\frac{z_{12} z_{34}}{z_{13} z_{24}}$, and $\overline{\eta}_2=\frac{\overline{z}_{12} \overline{z}_{34}}{\overline{z}_{13} \overline{z}_{24}}$,
where
$z_{ij}=z_i-z_j$ and $\overline{z}_{ij}=\overline{z}_i-\overline{z}_j$.
For the choice of $A=[0,+\infty)$,
the cross ratios have the simple form 
\be
\begin{split}
\eta_2=&\frac{-(l-t)+\sqrt{(l-t)^2+(\beta+\delta\,t)^2}}{2 \sqrt{(l-t)^2+(\beta+\delta\,t)^2}},\\
\overline{\eta}_2=&\frac{-(l+t)+\sqrt{(l+t)^2+(\beta+\delta\,t)^2}}{2 \sqrt{(l+t)^2+(\beta+\delta\,t)^2}}.\\
\end{split}
\ee
In the limit $t\to +\infty$, one has
\be
\label{zzbar_longTime}
\begin{split}
\eta_2(t\to \infty)=&\frac{1+\sqrt{1+\delta^2}}{2\sqrt{1+\delta^2}},\quad 
\overline{\eta}_2(t\to \infty)=\frac{-1+\sqrt{1+\delta^2}}{2\sqrt{1+\delta^2}}.
\end{split}
\ee
As shown in Fig.\ref{Fig:Ising},
if the time evolution is real, i.e., $\delta=0$, one can find that 
$\eta_2(t\to \infty)=1$ and $\overline{\eta}_2(t\to \infty)=0$, and therefore $\Delta S_A^{(2)}=\log d_\sigma=\frac{1}{2}\log (2)\simeq 0.3466$, as predicted in \cite{2014_Song}. If $\delta>0$, i.e., we have a complex time evolution, then the steady value of $\Delta S_A^{(2)}$ depends on the concrete value of $\delta$, which can be explicitly obtained based on \eqref{S2} and \eqref{zzbar_longTime}.

Next, let us consider a finite $A$ with $A=[0,l_A]$.
The procedure of calcalating $\Delta S_A^{(2)}$ is the same as the previous case of $A=[0,+\infty)$ except that the conformal map in \eqref{map:w_z2} is changed to 
\be
\frac{w}{w-l_A}=z^2.
\ee
The time evolution of $\Delta S_A^{(2)}$ is shown in Fig.\ref{Fig:Ising}. In the real time evolution with $\delta=0$, one can find that $\Delta S_A^{(2)}=\log d_\sigma$ in the region $t\in[l,l+l_A]$ and zero otherwise. In the complex time evolution with $\delta>0$, one can find that $\Delta S_A^{(2)}$ in the region $t\in[l,l+l_A]$ is suppressed by increasing $\delta$, and then gradually decays to zero in time.

\end{document}